\documentclass[11pt]{article}

\usepackage[margin=2cm, includefoot, footskip=30pt]{geometry}
\usepackage{setspace}
\usepackage[numbers]{natbib}

\usepackage[boxruled]{algorithm2e}
\usepackage{amssymb}
\usepackage{amsmath}
\usepackage{booktabs} 
\usepackage{color}
\usepackage{enumerate} 
\usepackage{graphicx}
\usepackage{hyperref}
\usepackage{listings}
\usepackage{mathptmx}
\usepackage{subfig}

\newlength{\imgwidth}
\definecolor{deepblue}{rgb}{0,0,0.5}
\definecolor{deepred}{rgb}{0.6,0,0}
\definecolor{deepgreen}{rgb}{0,0.5,0}

\newcommand{\balg}[1][htbp]{%
    \begin{algorithm}[#1]\DontPrintSemicolon
}
\newcommand{\ealg}{%
    \end{algorithm}
}

\renewcommand{\figurename}{Figure}
\DeclareMathOperator*{\argmin}{arg\,min}

\title{%
    Evolutionary Dataset Optimisation:
    learning algorithm quality through evolution
}
\author{Henry Wilde, Vincent Knight and Jonathan Gillard \\
    \textit{School of Mathematics, Senghennydd Rd, Cardiff, WALES CF24 4AG} \\
    \texttt{\{wildehd, knightva, gillardjw\}@cardiff.ac.uk}
}

\begin{document}
\bibliographystyle{spmpsci}

\maketitle%

\begin{abstract}
    In this paper we propose a novel method for learning how algorithms perform.
    Classically, algorithms are compared on a finite number of existing (or
    newly simulated) benchmark datasets based on some fixed metrics. The
    algorithm(s) with the smallest value of this metric are chosen to be the
    `best performing'. We offer a new approach to flip this paradigm. We
    instead aim to gain a richer picture of the performance of an algorithm by
    generating artificial data through genetic evolution, the purpose of which
    is to create populations of datasets for which a particular algorithm
    performs well on a given metric. These datasets can be studied so as to
    learn what attributes lead to a particular progression of a given algorithm.
    Following a detailed description of the algorithm as well as a brief
    description of an open source implementation, a case study in clustering is
    presented. This case study demonstrates the performance and nuances of the
    method which we call Evolutionary Dataset Optimisation. In this study, a
    number of known properties about preferable datasets for the clustering
    algorithms known as \(k\)-means and DBSCAN are realised in the generated 
    datasets.
\end{abstract}

\section{Introduction}\label{section:introduction}

This work presents a novel approach to learning the quality and performance of
an algorithm through the use of evolution. When an algorithm is developed to
solve a given problem, the designer is presented with questions about the
performance of their proposed method and its relative performance against
existing methods. This is an inherently difficult task. However, under the
current paradigm, the standard response to this situation is to use a known
fixed set of datasets \-- or simulate new datasets themselves \-- and a common
metric amongst the proposed method and its competitors. The collated algorithms
are then assessed based on this metric with often minimal consideration for
the appropriateness or reliability of the datasets being used, and the
robustness of the method(s) in question~\cite{Abualigah2018a,Huang1998,Liu2016}.

This process is not so readily observed when travelling in the opposite
direction but methods to do so exist. Suppose that the object of interest was
not an algorithm but rather a dataset. In this case, the objective is to
determine a preferable algorithm to complete some task on the data.  There exist
a number of methods employed across disciplines to complete this task that take
into account the characteristics of the data and the context of the research
problem. These methods are often equivalent to asking questions of the data, and
include the use of diagnostic tests. For instance, in the case of clustering, if
the data displayed an indeterminate number of non-convex blobs, then one could
recommend that an appropriate clustering algorithm would be
DBSCAN~\cite{Ester1996}. Otherwise, for scalability, \(k\)-means may be
chosen~\cite{Wu2009,Zhao2009}.

\begin{figure}[htbp]
    \centering
    \includegraphics{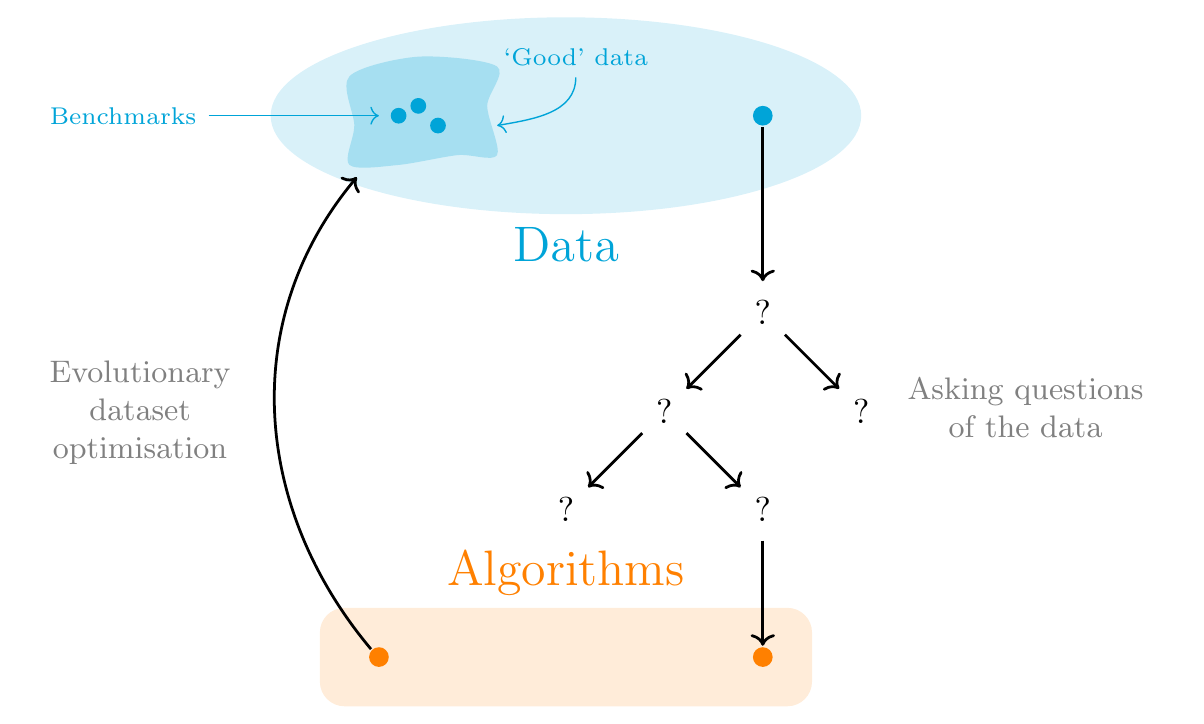}
    \caption{%
        On the right: the current path for selecting some algorithm(s) based on
        their validity and performance for a given dataset. On the left: the
        proposed flip to better understand the space in which `good' datasets
        exist for an algorithm.
    }\label{fig:paradigm}
\end{figure}

The approach presented in this work aims to flip the paradigm described here by
allowing the data itself to be unfixed. This fluidity in the data is achieved by
generating data for which the algorithm performs well (or better than some
other) through the use of an evolutionary algorithm. The purpose of doing so is
not to simply create a bank of useful datasets but rather to allow for the
subsequent studying of these datasets. In doing so, the attributes and
characteristics which lead to the success (or failure) of the algorithm may be
described, giving a broader understanding of the algorithm on the whole. Our
framework is described in Figure~\ref{fig:paradigm}.

This proposed flip has a number of motivations, and below is a non-exhaustive
list of some of the problems that are presented by the established evaluation
paradigm:
\begin{enumerate}
    \item How are these benchmark examples selected? There is no true measure of
        their reliability other than their frequent use. In some domains and
        disciplines there are well-established benchmarks so those found through
        literature may well be reliable, but in others less
        so~\cite{Campos2016,UCRArchive2018,Wang2015}.
    \item Sometimes, when there is a lack of benchmark examples, a `new' dataset
        is simulated to assess the algorithm~\cite{Olson2017}. This begs the
        question as to how and why that simulation is created. Not only this,
        but the origins of existing benchmarks is often a matter of convenience
        rather than their merit.
    \item In disciplines where there are established benchmarks, there may still
        be underlying problems around the true performance of an algorithm:
        \begin{enumerate}[(i)]
            \item As an example, work by Torralba and Efros~\cite{Torralba2011}
                showed that image classifiers trained and evaluated on a
                particular dataset, or datasets, did not perform reliably when
                evaluated using other benchmark datasets that were determined
                to be similar. Thus leading to a model which lacks robustness.
            \item The amount of learning one can gain as to the characteristics
                of data which lead to good (or bad) performance of an algorithm
                is constrained to the finite set of attributes present in the
                benchmark data chosen in the first place.
        \end{enumerate}
\end{enumerate}

\begin{figure}[htbp]
    \centering
    \includegraphics[width=\imgwidth]{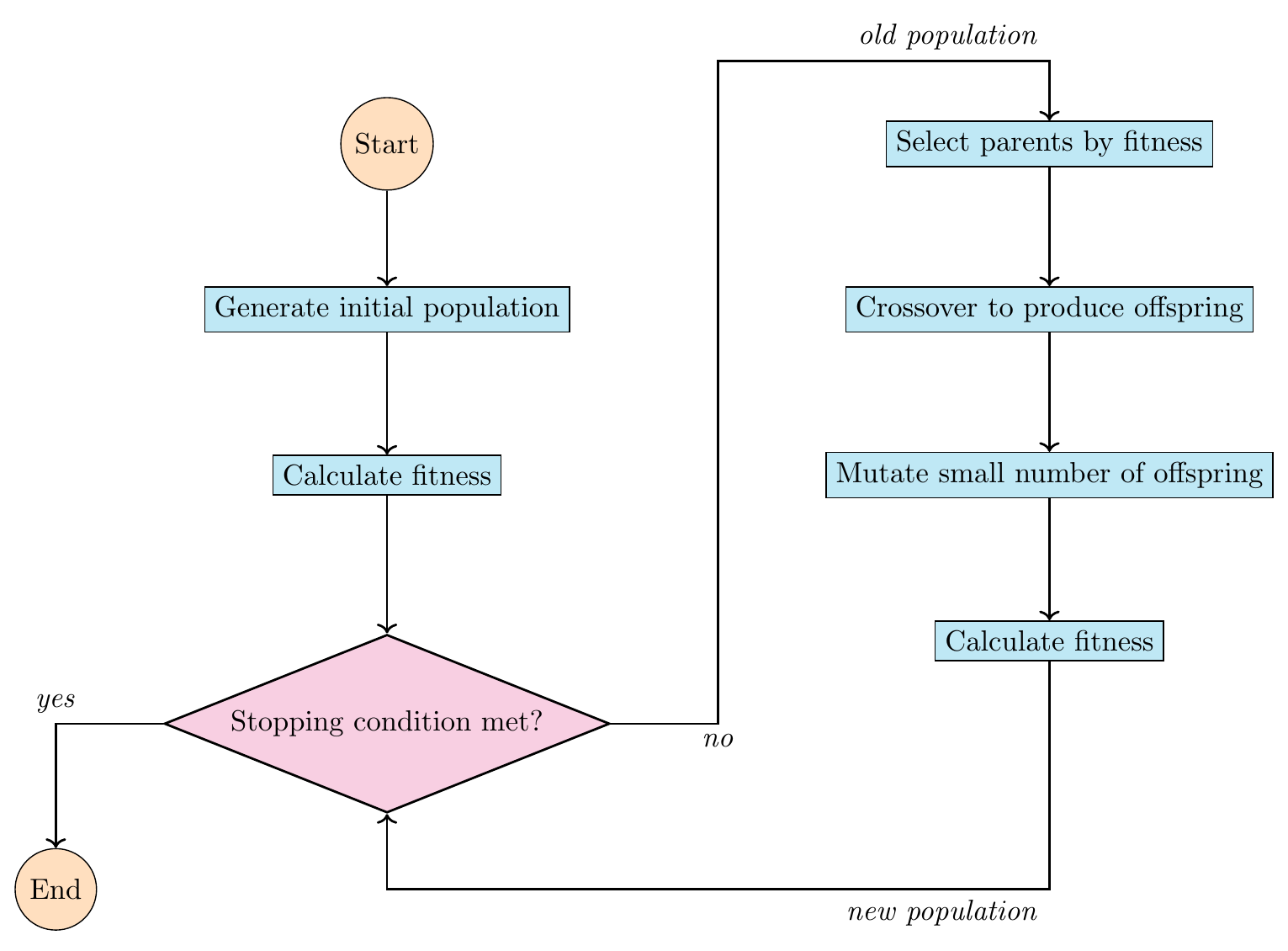}
    \caption{%
        A general schematic for an evolutionary algorithm.
    }\label{fig:flowchart}
\end{figure}

This work presents just one method from this new paradigm, and that method is
built around the concept of evolution. Evolutionary algorithms (EAs) have been
applied successfully to solve a wide array of problems \-- particularly where
the complexity of the problem or its domain are significant. These methods are
highly adaptive and their population-based construction (displayed in
Figure~\ref{fig:flowchart}) allows for the efficient solving of problems that
are otherwise beyond the scope of traditional search and optimisation methods.
EAs have been chosen here as they are simple in design yet their capabilities
encompass the difficulties of the flipped paradigm set out above.

The use of EAs to generate artificial data is not a new concept, however. Its
applications in data generation have included developing methods for the
automated testing of software~\cite{Koleejan2015,Michael2001,Sharifipour2018}
and the synthesis of existing or confidential data~\cite{Chen2016}. Such methods
also have a long history in the parameter optimisation of algorithms, and
recently in the automated design of convolutional neural network (CNN)
architecture~\cite{Suganuma2017,Sun2018}.

Other methods for the generation or synthesis of artificial data are numerous
and range from simple concepts such as simulated annealing~\cite{Matejka2017}
to swarm-based learning techniques~\cite{Abualigah2018b} or generative
adversarial networks (GANs)~\cite{Goodfellow2014}. The unconstrained learning
style of methods such as CNNs and GANs aligns with that proposed in this work.
By allowing the EA to explore and learn about the search space in an organic
way, less-prejudiced insight can be established that is not necessarily reliant
on any particular framework or agenda.

Note that the proposed methodology is not simply to use an EA to optimise an
algorithm over a search space with fixed dimension or data type such as those
set out in~\cite{Chen2016}. The shape of a dataset is considered a part of the
sample space itself that can be traversed through the evolutionary algorithm.

The remainder of the paper is structured as follows:
\begin{itemize}
    \item Section 2 describes the structure of the proposed method including its
        parameters and operators.
    \item Section 3 contains a case study where the success and failure of
        \(k\)-means clustering is examined using the proposed method. Included
        also is a comparison between \(k\)-means and another clustering
        algorithm DBSCAN.\
    \item Section 4 concludes this paper.
\end{itemize}

\section{The evolutionary algorithm}\label{section:algorithm}

In this section, the details of an algorithm that generates data for which a
given function, or (equivalently) algorithm, is well-suited is described. This
algorithm is to be referred to as ``Evolutionary Dataset Optimisation'' (EDO).

The EDO method is built as an evolutionary algorithm which follows a traditional
(generic) schema with some additional features that keep the objective of
artificial data generation in mind. With that, there are a number of parameters
that are passed to EDO;\ the typical parameters of an evolutionary algorithm
are a fitness function, \(f\), which maps from an individual to a real number,
as well as a population size, \(N\), a maximum number of iterations, \(M\), a
selection parameter, \(b\), and a  mutation probability, \(p_m\). In addition to
these, EDO takes the following parameters:
\begin{itemize}
    \item A set of probability distribution families, \(\mathcal{P}\). Each
        family in this set has some parameter limits which form a part of the
        overall search space. For instance, the family of normal distributions,
        denoted by \(N(\mu, \sigma^2)\), would have limits on values for the
        mean, \(\mu\), and the standard deviation, \(\sigma\).
    \item A maximum number of ``subtypes'' for each family in \(\mathcal{P}\). A
        subtype is an independent copy of the family that progresses separate
        from the others. These are the actual distribution objects which are
        traversed in the optimisation.
    \item A probability vector to sample distributions from \(\mathcal{P}\),
        \(w = \left(w_1, \ldots, w_{|\mathcal{P}|}\right)\).
    \item Limits on the number of rows an individual dataset can have,
        \[
            R \in \left\{%
                (r_{\min}, r_{\max}) \in \mathbb{N}^2~|~r_{\min} \leq r_{\max}
            \right\}
        \]
    \item Limits on the number of columns a dataset can have,
        \[
            C := \left(C_1, \ldots, C_{|\mathcal{P}|}\right)
            \text{ where }
            C_j \in \left\{ (c_{\min}, c_{\max}) \in {%
                \left(\mathbb{N}\cup\{\infty\}\right)
            }^2~|~c_{\min} \leq c_{\max}\right\}
        \]
        for each \(j = 1, \ldots, |\mathcal{P}|\). That is, \(C\) defines the
        minimum and maximum number of columns a dataset may have from each
        distribution in \(\mathcal{P}\).
    \item A second selection parameter, \(l \in [0, 1]\), to allow for a
        small proportion of `lucky' individuals to be carried forward.
    \item A shrink factor, \(s \in [0, 1]\), defining the relative size of a
        component of the search space to be retained after adjustment.
\end{itemize}

The concepts discussed in this section form the mechanisms of the evolutionary
dataset optimisation algorithm. To use the algorithm practically, these
components have been implemented in Python as a library built on the scientific
Python stack~\cite{pandas,numpy}. The library is fully tested and documented (at
\url{https://edo.readthedocs.io}) and is freely available online under the MIT
license~\cite{edo-project}. The EDO implementation was developed to be
consistent with the current best practices of open source software
development~\cite{Jiminez2017}.

\balg%
\KwIn{\(f, N, R, C, \mathcal{P}, w, M, b, l, p_m, s\)}
\KwOut{A full history of the populations and their fitnesses.}

\Begin{%
    create initial population of individuals\;
    find fitness of each individual\;
    record population and its fitness\;

    \While{%
        current iteration less than the maximum
        \textbf{and} stopping condition not met
    }{%
        select parents based on fitness and selection proportions\;
        use parents to create new population through crossover and mutation\;
        find fitness of each individual\;
        update population and fitness histories\;
        \If{adjusting the mutation probability}{%
            update mutation probability
        }
        \If{using a shrink factor}{%
            shrink the mutation space based on parents
        }
    }
}
\caption{The Evolutionary Dataset Optimisation algorithm}
\ealg\label{alg:edo}

\balg%
\KwIn{parents, \(N, R, C, \mathcal{P}, w, p_m\)}
\KwOut{A new population of size \(N\)}

\Begin{%
    add parents to the new population\;
    \While{the size of the new population is less than \(N\)}{%
        sample two parents at random\;
        create an offspring by crossing over the two parents\;
        mutate the offspring according to the mutation probability\;
        add the mutated offspring to the population\;
    }
}
\caption{Creating a new population}
\ealg%

The statement of the EDO algorithm is presented here to lay out its general
structure from a high level perspective. Lower level discussion is provided
below where additional algorithms for the individual creation, evolutionary
operator and shrinkage processes are given along with diagrams (where
appropriate). Note that there are no defined processes for how to stop the
algorithm or adjust the mutation probability, \(p_m\). This is down to their
relevance to a particular use case. Some examples include:
\begin{itemize}
    \item Regular decreasing in mutation probability across the available
        attributes~\cite{Kuehn2013}.
    \item Stopping when no improvement in the best fitness is found within some
        \(K\) consecutive iterations~\cite{Leung2001}.
    \item Utilising global behaviours in fitness to indicate a stopping
        point~\cite{Marti2016}.
\end{itemize}

\subsection{Individuals}

Evolutionary algorithms operate in an iterative process on populations of
individuals that each represent a solution to the problem in question. In a
genetic algorithm, an individual is a solution encoded as a bit string of,
typically, fixed length and treated as a chromosome-like object to be
manipulated. In EDO, as the objective is to generate datasets and explore the
space in which datasets exist, there is no encoding. As such the distinction is
made that EDO is an evolutionary algorithm. 

\begin{figure}[htbp]
    \centering
    \includegraphics[width=\imgwidth]{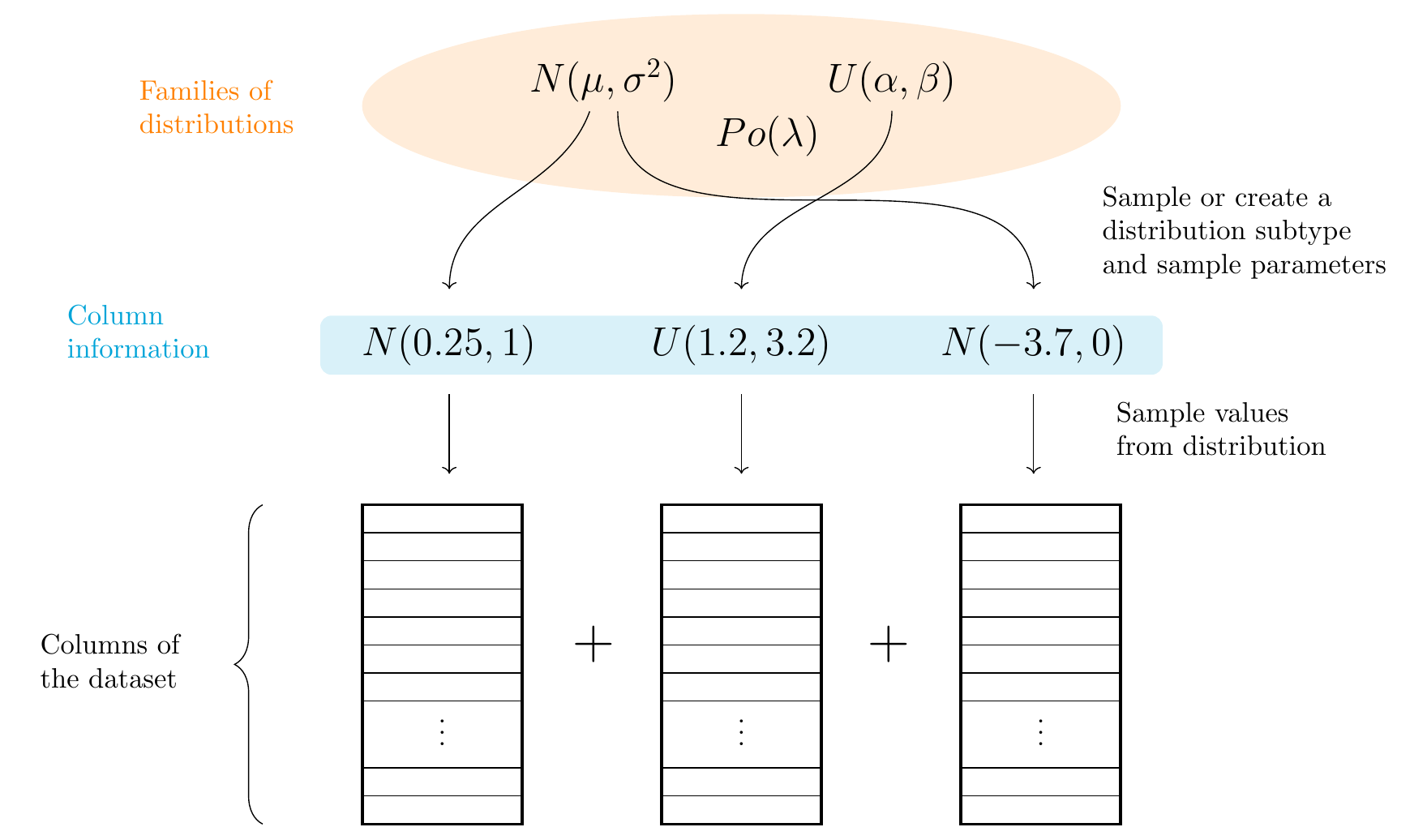}
    \caption{%
        An example of how an individual is first created.
    }\label{fig:individual}
\end{figure}

As is seen in Figure~\ref{fig:individual}, an individual's creation is
defined by the generation of its columns. A set of instructions on how to sample
new values (in mutation, for instance, Section~\ref{subsection:mutation}) for
that column are recorded in the form of a probability distribution. These
distributions are sampled and created from the families passed in
\(\mathcal{P}\). In EDO, the produced datasets and their metadata are
manipulated directly so that the biological operators can be designed and be
interpreted in a more meaningful way as will be seen later in this section.

However, one should not assume that the columns are a reliable representative of
the distribution associated with them, or vice versa. This is particularly true
of `shorter' datasets with a small number of rows, whereas confidence in the
pair could be given more liberally for `longer' datasets with a larger number
of rows. In any case, appropriate methods for analysis should be employed before
formal conclusions are made.

\balg%
\KwIn{\(R, C, \mathcal{P}, w\)}
\KwOut{An individual defined by a dataset and some metadata}

\Begin{%
    sample a number of rows and columns\;
    create an empty dataset\;
    \For{each column in the dataset}{%
        sample a distribution from \(\mathcal{P}\)\;
        create an instance of the distribution\;
        fill in the column by sampling from this instance\;
        record the instance in the metadata
    }
}
\caption{Creating an individual}
\ealg%

\subsection{Selection}

The selection operator describes the process by which individuals are chosen
from the current population to generate the next. Almost always, the likelihood
of an individual being selected is determined by their fitness. This is because
the purpose of selection is to preserve favourable qualities and encourage some
homogeneity within future generations~\cite{Back1994}.

\begin{figure}[htbp]
    \centering
    \includegraphics[width=.8\imgwidth]{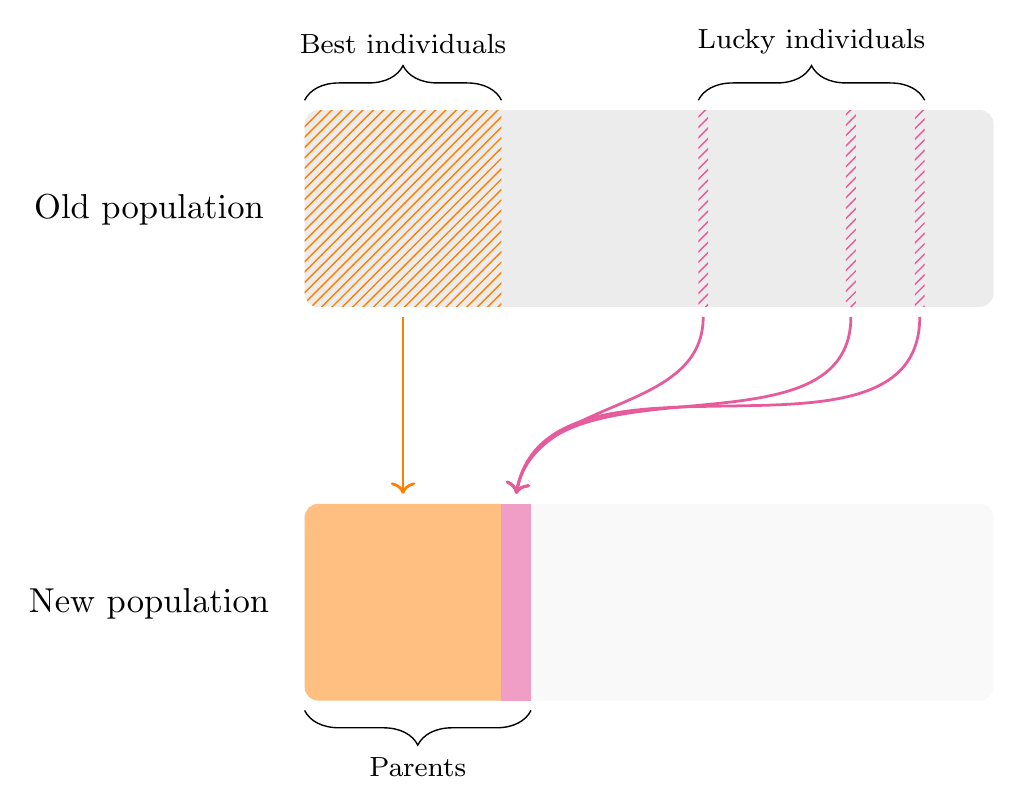}
    \caption{%
        The selection process with the inclusion of some lucky individuals.
    }\label{fig:selection}
\end{figure}

\balg%
\KwIn{population, population fitness, \(b\), \(l\)}
\KwOut{A set of parent individuals}

\Begin{%
    calculate \(n_b\) and \(n_l\)\;
    sort the population by the fitness of its individuals\;
    take the first \(n_b\) individuals and make them parents\;
    \If{there are any individuals left}{%
        take the next \(n_l\) individuals and make them parents\;
    }
}
\caption{The selection process}
\ealg%

In EDO, a modified truncation selection method is used~\cite{Jebari2013}, as can
be seen in Figure~\ref{fig:selection}. Truncation selection takes a fixed
number, \(n_b = \lceil bN\rceil\), of the fittest individuals in a population
and makes them the `parents' of the next. It has been observed that, despite
its efficiency as a selection operator, truncation selection can lead to
premature convergence at local optima~\cite{Jebari2013,Motoki2002}. The
modification for EDO is an optional stage after the best individuals have been
chosen: with some small \(l\), a number, \(n_l = \lceil lN\rceil\), of the
remaining individuals can be selected at random to be carried forward. Hence,
allowing for a small number of randomly selected individuals may encourage
diversity and further exploration throughout the run of the algorithm. It should
be noted that regardless of this step, an individual could potentially be
present throughout the entirety of the algorithm.

After the parents have been selected, there are two adjustments made to the
current search space. The first is that the subtypes for each family in
\(\mathcal{P}\) are updated to only those present in the parents. The second
adjustment is a process which acts on the distribution parameter limits for
each subtype in \(\mathcal{P}\). This adjustment gives the ability to `shrink'
the search space about the region observed in a given population. This method is
based on a power law described in~\cite{Amirjanov2016} that relies on a shrink
factor, \(s\). At each iteration, \(t\), every distribution subtype which is
present in the parents has its parameter's limits, \(\left(l_t, u_t\right)\),
adjusted. This adjustment is such that the new limits, \(\left(l_{t+1},
u_{t+1}\right)\) are centred about the mean observed value, \(\mu\), for that
parameter:
\begin{align}
    \label{eq:shrinking_lower}
    l_{t+1}&= \max \left\{l_t, \ \mu - \frac{1}{2} (u_t - l_t) s^t\right\}\\
    \label{eq:shrinking_upper}
    u_{t+1}&= \min \left\{u_t, \ \mu + \frac{1}{2} (u_t - l_t) s^t\right\}
\end{align}

The shrinking process is given explicitly in
Algorithm~\ref{algorithm:shrinking}. Note that the behaviour of this process can 
produce reductive results for some use cases and is optional.

\balg%
\KwIn{parents, current iteration, \(\mathcal{P}, M, s\)}
\KwOut{A new mutation space focussed around the parents}

\Begin{%
    \For{each distribution subtype in \(\mathcal{P}\)}{%
        \For{each parameter of the distribution}{%
            get the current values for parameter over all parent columns\;
            find the mean of the current values\;
            find the new lower~(\ref{eq:shrinking_lower}) and
            upper~(\ref{eq:shrinking_upper}) bounds around the mean\;
            set the parameter limits\;
        }
    }
}
\caption{Shrinking the mutation space}\label{algorithm:shrinking}
\ealg%

\subsection{Crossover}

\begin{figure}[htbp]
    \centering
    \includegraphics[width=\linewidth]{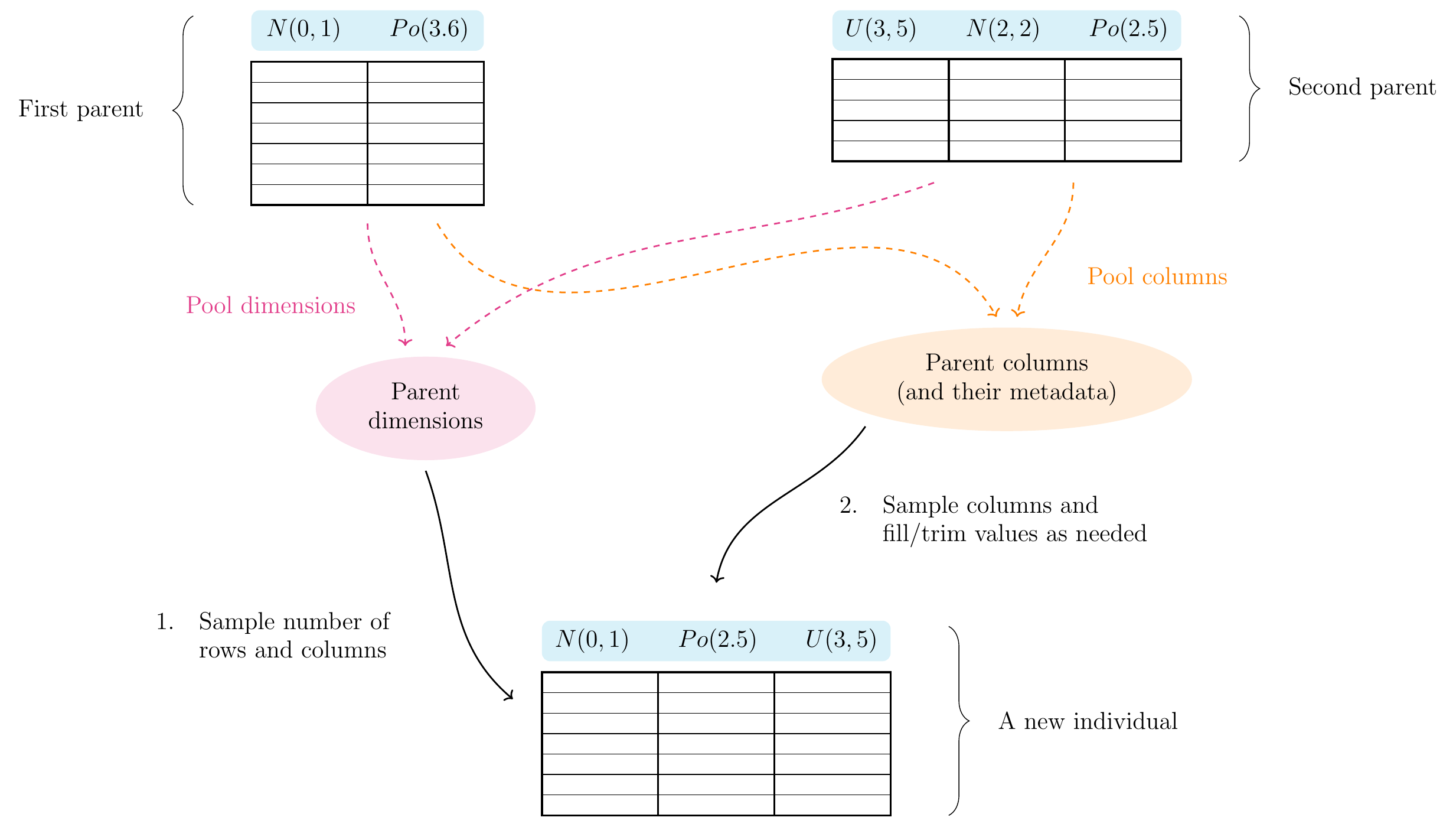}
    \caption{%
        The crossover process between two individuals with different dimensions.
    }\label{fig:crossover}
\end{figure}

Crossover is the operation of combining two individuals in order to create at
least one offspring. In genetic algorithms, the term `crossover' can be taken
literally: two bit strings are crossed at a point to create two new bit strings.
Another popular method is uniform crossover, which has been favoured for its
efficiency and efficacy in combining individuals in both bit string and matrix
representations~\cite{Chen2018,Semenkin2012}. For EDO, this method is adapted to
support dataset manipulation: a new individual is created by uniformly sampling
each of its components (dimensions and then columns) from a set of two `parent'
individuals, as shown in Figure~\ref{fig:crossover}.

Observe that there is no requirement on the dimensions of the parents to be of
similar or equal shapes. This is because the driving aim of the proposed method
is to explore the space of all possible datasets. In the case where there is
incongruence in the lengths of the two parents, missing values may appear in a
shorter column that is sampled. To resolve this, values are sampled from the
probability distribution associated with that column to fill in these gaps.

\balg%
\KwIn{Two parents}
\KwOut{An offspring made from the parents ready for mutation}

\Begin{%
    collate the columns and metadata from each parent in a pool\;
    sample each dimension from between the parents uniformly\;
    form an empty dataset with these dimensions\;
    \For{each column in the dataset}{%
        sample a column (and its corresponding metadata) from the pool\;
        \If{this column is longer than required}{%
            randomly select entries and delete them as needed 
        }
        \If{this column is shorter than required}{%
            sample new values from the metadata and append them to the column as
            needed
        }
        add this column to the dataset and record its metadata\;
    }
}
\caption{The crossover process}
\ealg%

\subsection{Mutation}\label{subsection:mutation}

Mutation is used in evolutionary algorithms to encourage a broader exploration
of the search space at each generation. Under this framework, the mutation
process manipulates the phenotype of an individual where numerous things need to
be modified including an individual's dimensions, column metadata and the
entries themselves. This process is described in Figure~\ref{fig:mutation}.

\begin{figure}[htbp]
    \centering
    \includegraphics[width=\linewidth]{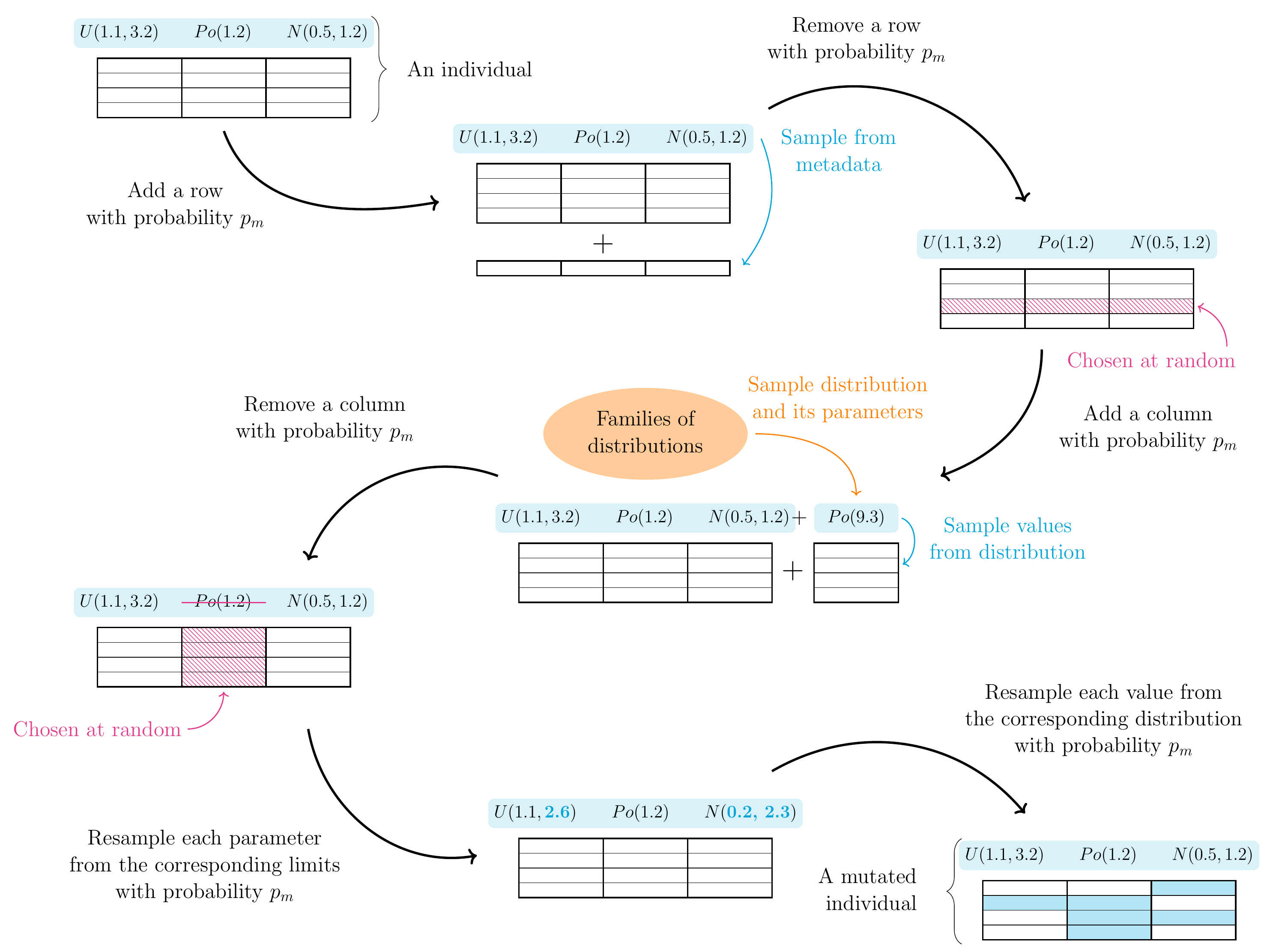}
    \caption{%
        The stages of the mutation process.
    }\label{fig:mutation}
\end{figure}

As shown in Figure~\ref{fig:mutation}, each of the potential mutations occur
with the same probability, \(p_m\). However, the way in which columns are
maintained assure that (assuming appropriate choices for \(f\) and
\(\mathcal{P}\)) many mutations in the metadata and the dataset itself will only
result in some incremental change in the individual's fitness
relative to, say, a completely new individual.

\balg%
\KwIn{An individual, \(p_m\), \(R\), \(C\), \(\mathcal{P}\), \(w\)}
\KwOut{A mutated individual}

\Begin{%
    sample a random number \(r \in [0, 1]\)\;
    \If{\(r < p_m\) and adding a row would not violate \(R\)}{%
        sample a value from each distribution in the metadata\;
        append these values as a row to the end of the dataset\;
    }
    sample a new \(r \in [0, 1]\)\;
    \If{\(r < p_m\) and removing a row would not violate \(R\)}{%
        remove a row at random from the dataset
    }
    sample a new \(r \in [0, 1]\)\;
    \If{\(r < p_m\) and adding a new column would not violate \(C\)}{%
        create a new column using \(\mathcal{P}\) and \(w\)\;
        append this column to the end of the dataset
    }
    sample a new \(r \in [0, 1]\)\;
    \If{\(r < p_m\) and removing a column would not violate \(C\)}{%
        remove a column (and its associated metadata) at random from the dataset
    }
    \For{each distribution in the metadata}{%
        \For{each parameter of the distribution}{%
            sample a random number \(r \in [0, 1]\)\;
            \If{\(r < p_m\)}{%
                sample a new value from within the distribution parameter
                limits\;
                update the parameter value with this new value
            }
        }
    }
    \For{each entry in the dataset}{%
        sample a random number \(r \in [0, 1]\)\;
        \If{\(r < p_m\)}{%
            sample a new value from the associated column distribution\;
            update the entry with this new value
        }
    }
}
\caption{The mutation process}\label{algorithm:mutation}
\ealg%

\section{A case study in clustering}\label{section:examples}

\subsection{\(k\)-means clustering}

The following examples act as a form of validation for EDO, and also highlight
some of the nuances in its use. The objective of these examples is to use the
proposed method to reproduce some known results about the clustering of data in
the absence of any external forces, and to examine how clustering algorithms are
typically evaluated. In particular, the focus will be on the well-known
\(k\)-means (Lloyd's) algorithm. Clustering was chosen as it is a
well-understood problem that is easily accessible \-- especially when restricted
to two dimensions. The \(k\)-means algorithm is an iterative, centroid-based
method that aims to minimise the `inertia' of the current partition, \(Z =
\left\{Z_1, \ldots, Z_k\right\}\), of some dataset \(X\):
\begin{equation}
    I(Z, X) := \frac{1}{|X|} \sum_{j=1}^{k} \sum_{x \in Z_j} {d(x, z_j)}^2
    \label{eq:inertia}
\end{equation}

A full statement of the algorithm to minimise~(\ref{eq:inertia}) is given
in~\ref{app:kmeans}. 

This inertia function is taken as the objective of the \(k\)-means algorithm,
and is used for evaluating the final clustering. This is particularly true when
the algorithm is not being considered an unsupervised classifier where accuracy
may be used~\cite{Huang1998}. With that, the first example will use this inertia
as the fitness function in EDO.\ That is, to find datasets which minimise \(I\).

For the purposes of visualisation, EDO is restricted to the space containing
only two-dimensional datasets, i.e.\ \(C = \left((2, 2)\right)\). In addition to
this, all columns are formed from uniform distributions where the bounds are
sampled from the unit interval. Thus, the only family in \(\mathcal{P}\) is:
\begin{equation}
    \mathcal{U} := \left\{U(a, b)~|~a, b \in [0, 1]\right\}
\end{equation}

The remaining parameters are as follows: \(N~=~100\), \(R~=~(3, 100)\),
\(M~=~1000\), \(b~=~0.2\), \(l~=~0\), \(p_m~=~0.01\), and shrinkage is excluded.
Figure~\ref{fig:small-inertia-50} shows an example of the fitness (above) and
dimension (below) progression of the evolutionary algorithm under these
conditions up until the \(50^{th}\) epoch.

There is a steep learning curve here; within the first 50 generations an
individual is found with a fitness of roughly \(10^{-10}\) which could not be
improved on for a further 900 epochs. The same quick convergence is seen in the
number of rows. This behaviour is quickly recognised as preferable and was
dominant across all the trials conducted in this work. This preference for
datasets with fewer rows is expected given that \(I\) is the sum of the mean
error from each cluster centre. With that, when \(k\) is fixed \textit{a
priori}, reducing the number of points in each cluster (i.e.\ the terms of the
second summation) quickly reduces the mean error of that cluster and thus the
value of \(I\).

\addtocounter{figure}{1}
\begin{figure}[htbp]
    \ContinuedFloat%
    \centering
    \begin{tabular}{c}
        \includegraphics[width=\imgwidth]{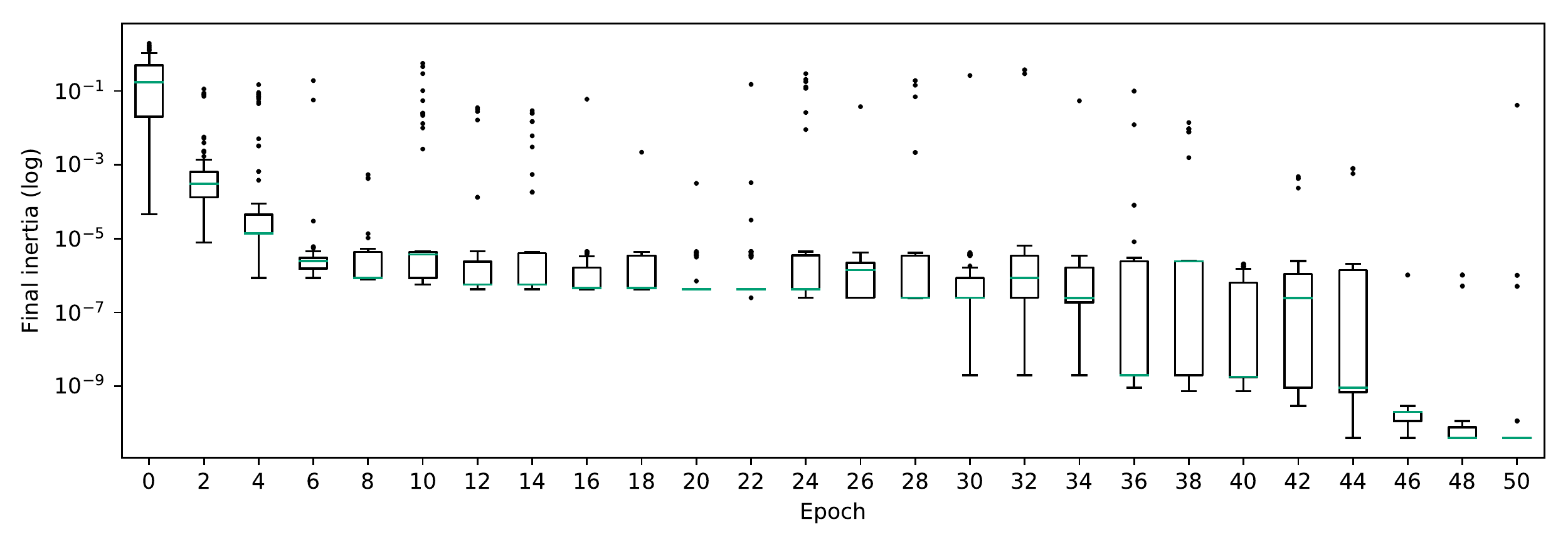}
        \\
        \includegraphics[width=\imgwidth]{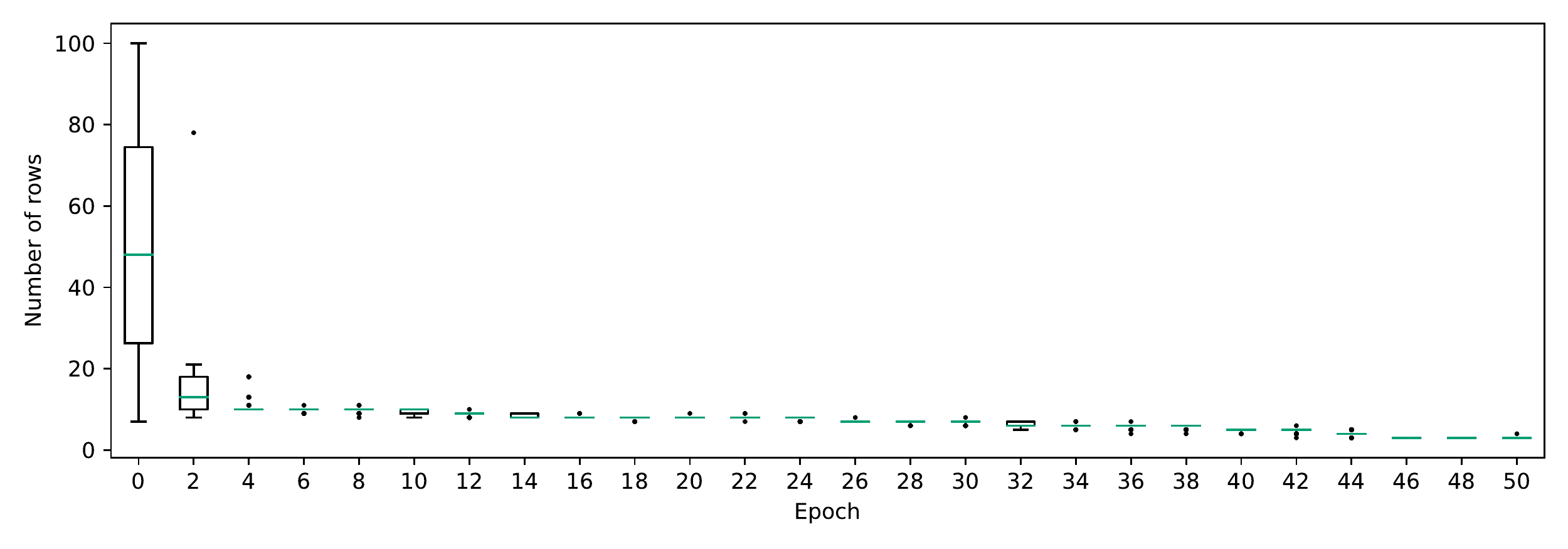}
    \end{tabular}
    \caption{%
        Progressions for final inertia and dimension across the first 50
        epochs with \(R~=~(3,100)\).
    }\label{fig:small-inertia-50}
\end{figure}

\begin{figure}[htbp]
    \ContinuedFloat%
    \centering
    \begin{tabular}{c}
        \includegraphics[width=\imgwidth]{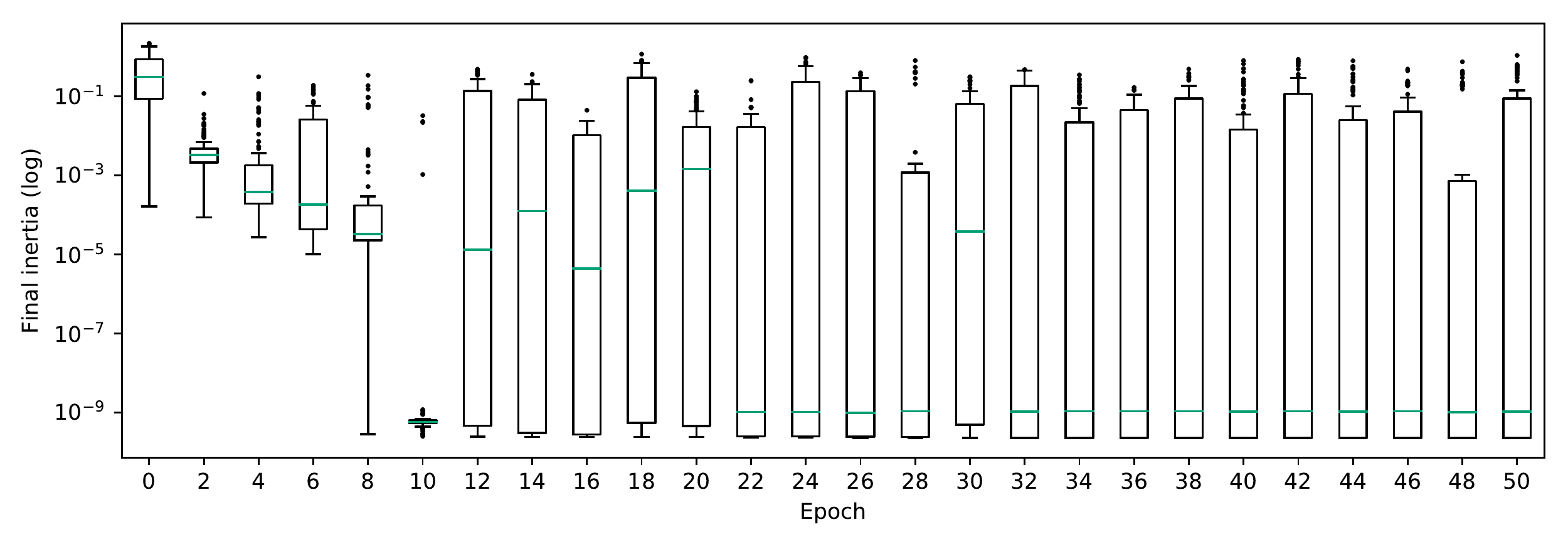}
        \\
        \includegraphics[width=\imgwidth]{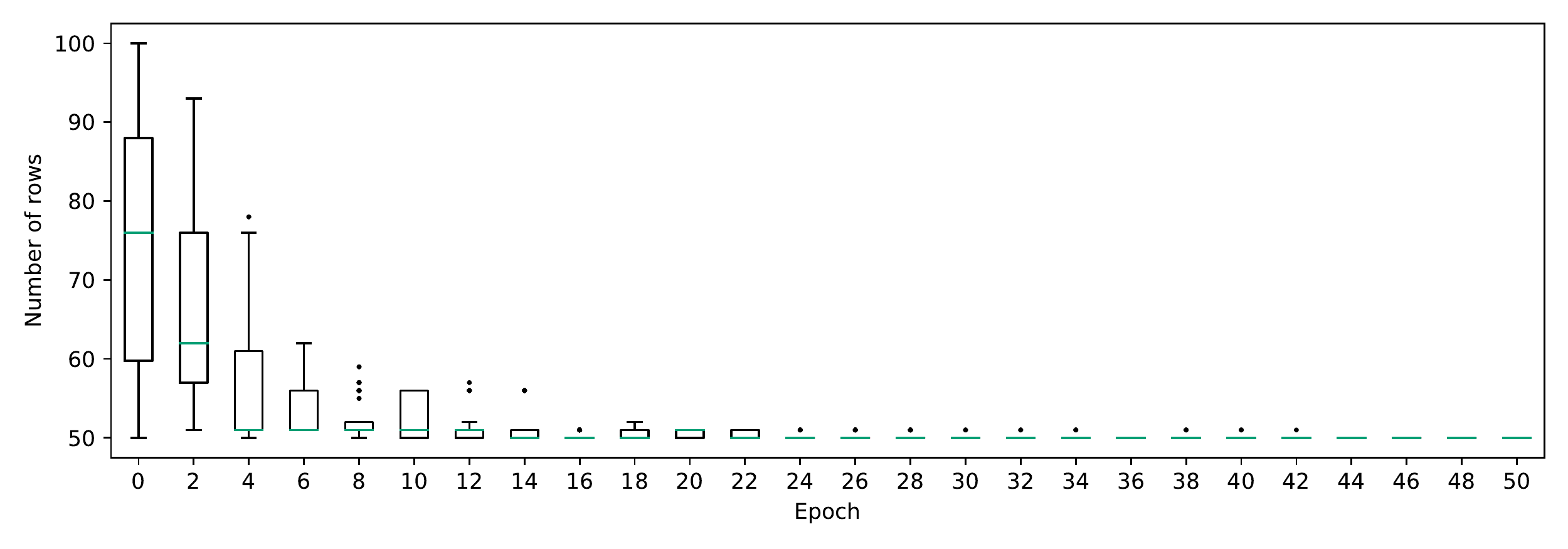}
    \end{tabular}
    \caption{%
        Progressions for final inertia and dimension across the first 50 epochs
        with \(R~=~(50,100)\).
    }\label{fig:large-inertia-50}
\end{figure}

However, something that may be seen as unwanted is a compaction of the cluster
centres. Referring to Figure~\ref{fig:small-inertia-inds}, the best and median
individuals show two clusters that are essentially the same point whereas the
worst is a random cloud across the whole of \(\mathcal{U}\) which was found in
the initial population. The kind of behaviour exhibited by the best performing
individuals here occurs in part because it is allowed. There are two immediate
ways in which this allowed: first, that a near-trivial case is included in \(R\)
and, secondly, that the fitness function does nothing to penalise the proximity
of the inter-cluster means, as well as aiming to reduce the intra-cluster means.
This kind of unwanted behaviour highlights a subtlety in how EDO should be used;
that experimentation and rigour are required to properly understand an
algorithm's quality.

\begin{figure}[htbp]
    \centering
    \subfloat[][]{%
        \label{fig:small-inertia-inds}
        \centering
        \includegraphics[width=\imgwidth]{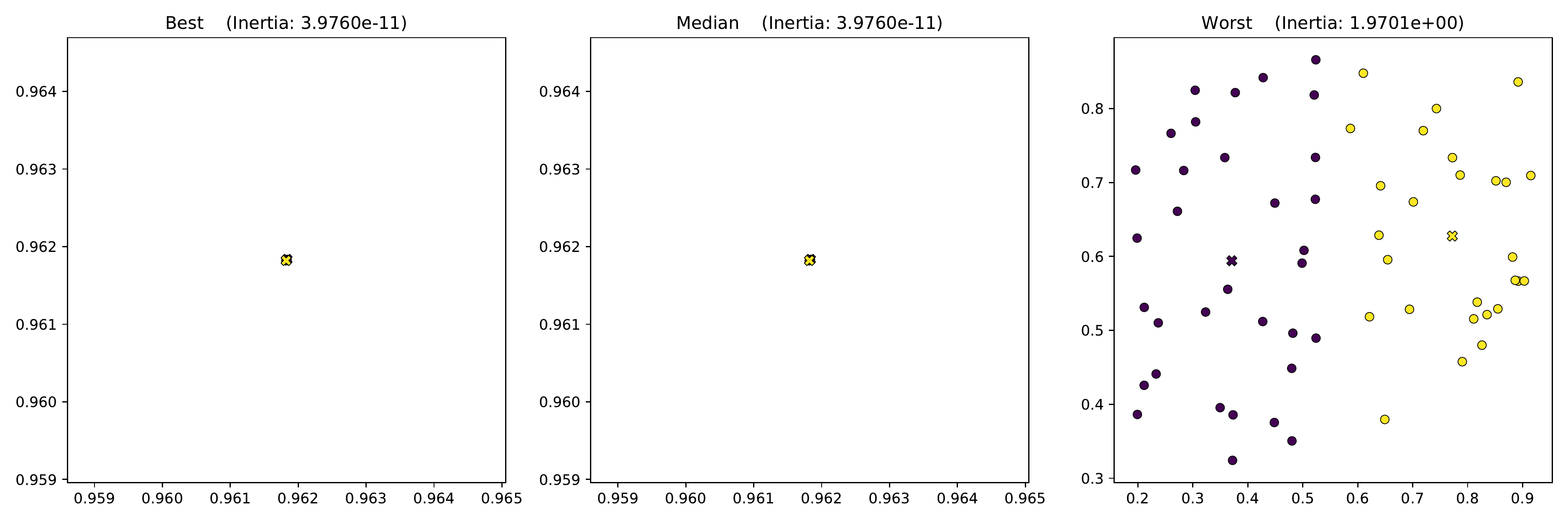}
    }\\

    \subfloat[][]{%
        \label{fig:large-inertia-inds}
        \centering
        \includegraphics[width=\imgwidth]{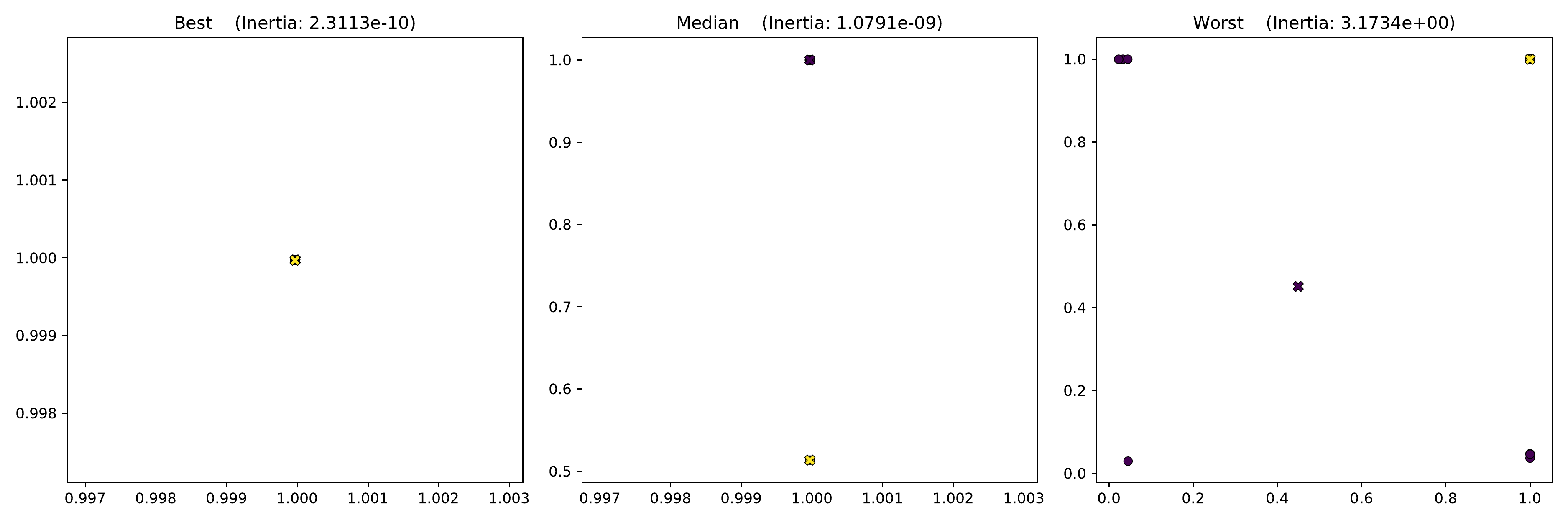}
    }
    \caption[]{%
        Representative individuals based on inertia with:
        \subref{fig:small-inertia-inds} \(R~=~(3,100)\);
        \subref{fig:large-inertia-inds} \(R~=~(50,100)\). Centroids displayed as
        crosses.
    }\label{fig:inertia-inds}
\end{figure}

Hence, consider Figure~\ref{fig:large-inertia-inds} where the individuals have
been generated with the same parameters as previously except with adjusted row
limits, \(R = (50, 100)\), so as to exclude this trivial case. In these trials,
the results are equivalent: the worst performing individuals are without
structure whilst the best-performing individuals display clusters that are dense
about a single point despite the minimum number of rows being increased.
Supposing this was not already a known result, we can see mounting evidence in
favour of this compaction being `optimal' behaviour in a dataset for \(k\)-means
clustering.

However, the fitness function may be addressed still, and more extensive
studying may be done. Indeed, the final inertia could be considered a flawed or
fragile fitness function if it is supposed to evaluate the efficacy of the
\(k\)-means algorithm. Incorporating the inter-cluster spread to the fitness
of an individual dataset would reduce this observed compaction. For instance,
the silhouette coefficient is a metric used to evaluate the appropriateness of a
clustering to a dataset and does precisely that. The silhouette coefficient of a
clustering of a dataset is given by the mean of the silhouette value,
\(S(x)\), of each point \(x \in Z_j\) in each cluster:
\begin{equation}
    \begin{gathered}
        A(x) := \frac{1}{|Z_j| - 1} \sum_{y \in Z_j \setminus \{x\}} d(x, y),
        \\
        B(x) := \min_{k \neq j} \frac{1}{|Z_k|} \sum_{w \in Z_k} d(x, w),
        \\
        S(x) := 
            \begin{cases}
                \frac{B(x) - A(x)}{\max\left\{A(x), B(x)\right\}}
                &\quad \text{if } |Z_j| > 1\\
                0 &\quad \text{otherwise}
            \end{cases}
    \end{gathered}\label{eq:silhouette}
\end{equation}\\

The optimisation of the silhouette coefficient is analogous to finding a dataset
which increases both the intra-cluster cohesion (the inverse of \(A\)) and
inter-cluster separation (\(B\)). Hence, the objective of minimising inertia is
addressed by maximising cohesion. Meanwhile, the additional desire to spread out
the clusters is considered by maximising separation.

Repeating the trials with the same parameters as with inertia, the silhouette
fitness function yields the results summarised in
Figures~\ref{fig:small-silhouette}~and~\ref{fig:large-silhouette}. Irrespective
of row limits, the datasets produced show increased separation from one another
whilst maintaining low values in the final inertia of the clustering as shown in
Figure~\ref{fig:silhouette-inds}. Again, the form of the individual clusters is
much the same. The low values of inertia correspond to tight clusters, and the
tightest clusters are those with a minimal number of points, i.e.\ a single
point. As with the previous example, albeit at a much slower rate, the
preferable individuals are those leading toward this case. That this gradual
reduction in the dimension of the individuals occurs despite adjusting the
fitness function and considering the space which excludes the trivial case
bolsters the claim that the base case is also optimal.

At this point, it should be noted that, due to the nature of the implementation,
any individual from any generation may be retrieved and studied should the final
results be too concentrated on any given case. The summary provided here is one
particular way of studying the body of datasets generated with this method and
this transparency in the history and progression of the proposed method is
something that sets it apart from other methods such as GANs which have a
reputation of providing so-called `black box' solutions.

\addtocounter{figure}{1}
\begin{figure}[htbp]
    \ContinuedFloat%
    \centering
    \begin{tabular}{c}
        \includegraphics[width=\imgwidth]{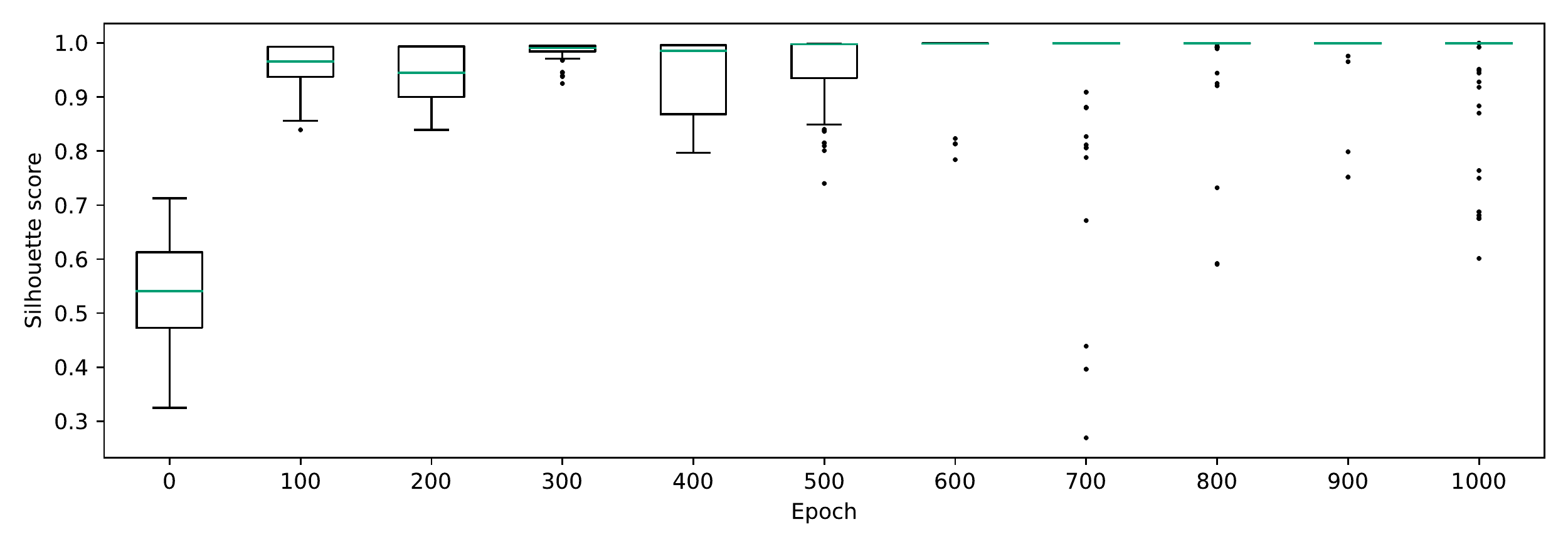}
        \\
        \includegraphics[width=\imgwidth]{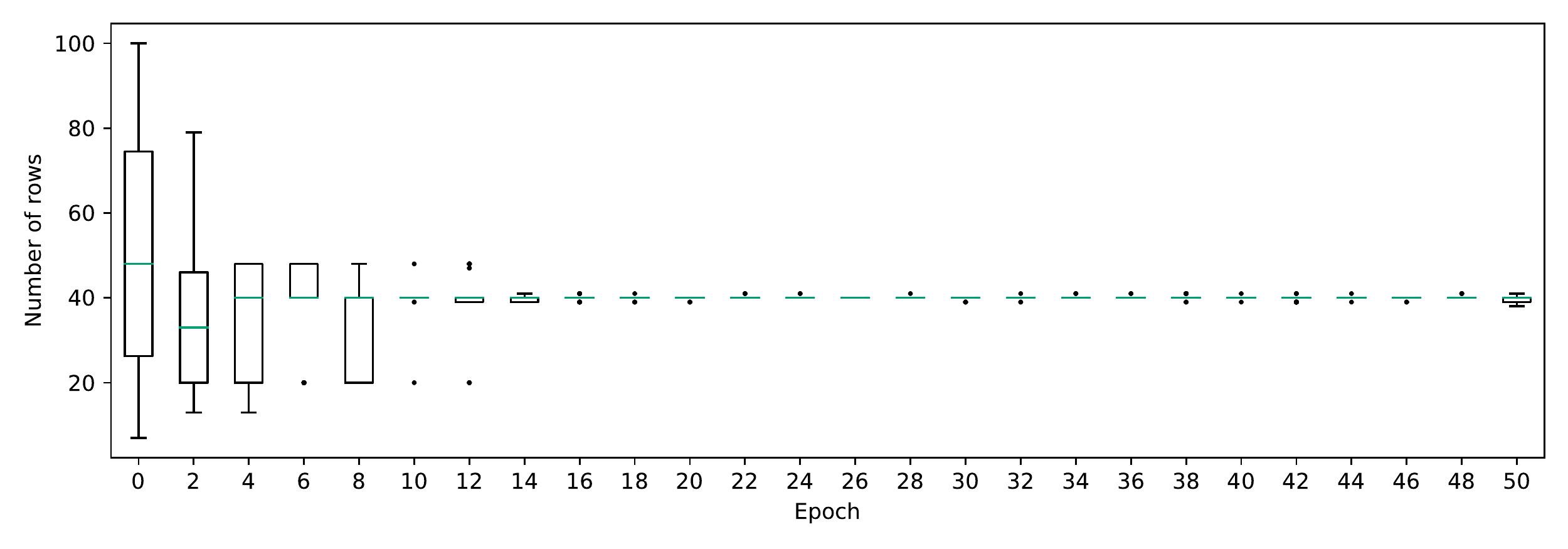}
    \end{tabular}
    \caption{%
        Progressions for silhouette and dimension across 1000 epochs at 100
        epoch intervals with \(R~=~(3, 100)\).
    }\label{fig:small-silhouette}
\end{figure}

\begin{figure}
    \ContinuedFloat%
    \centering
    \begin{tabular}{c}
        \includegraphics[width=\imgwidth]{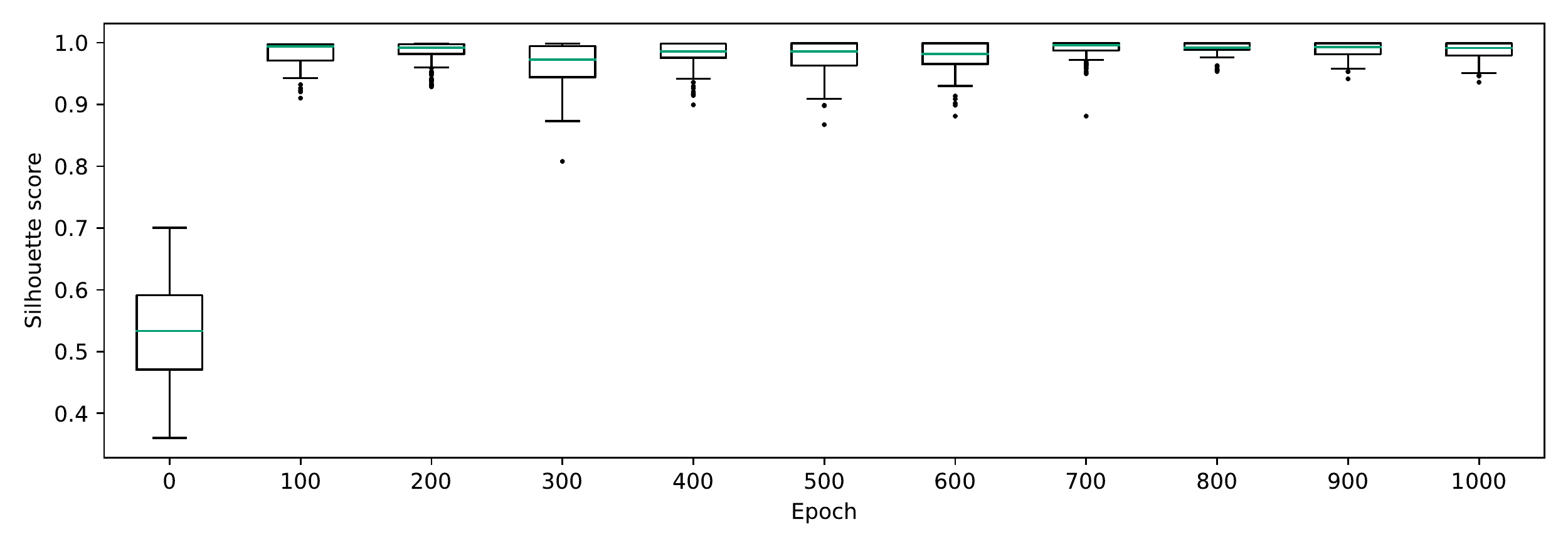}
        \\
        \includegraphics[width=\imgwidth]{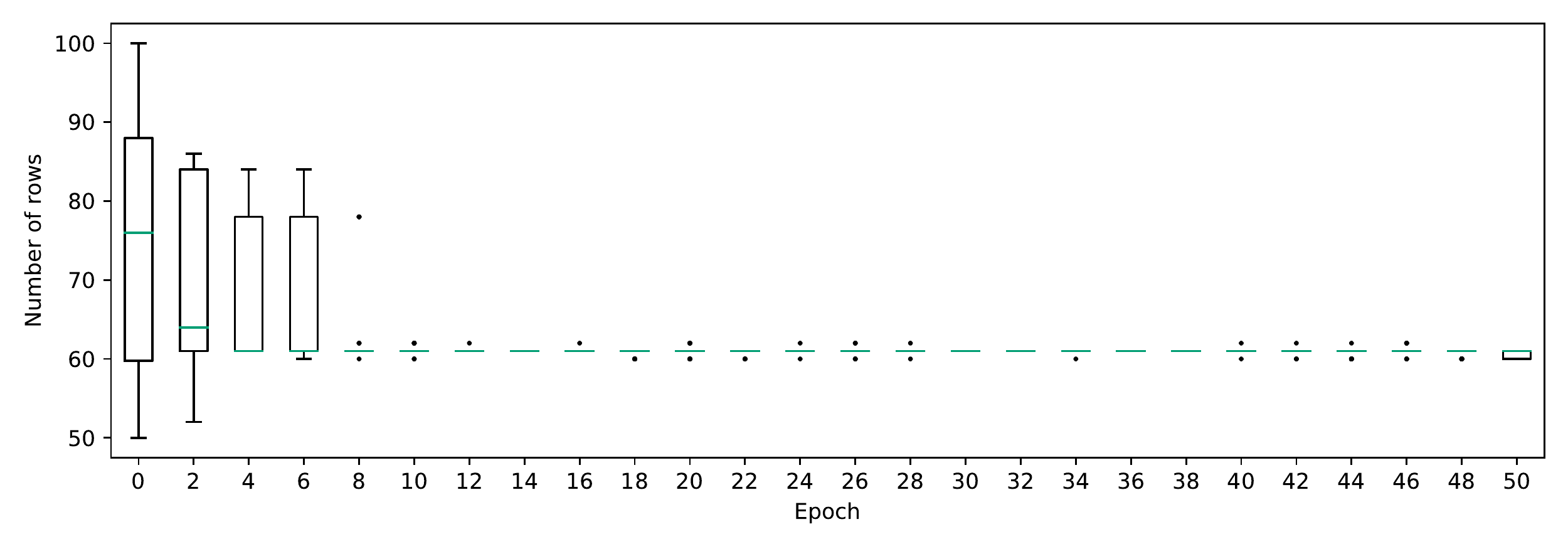}
    \end{tabular}
    \caption{%
        Progressions for silhouette and dimension across 1000 epochs at 100
        epoch intervals with \(R~=~(50,100)\).
    }\label{fig:large-silhouette}
\end{figure}

\begin{figure}[htbp]
    \centering
    \subfloat[][]{%
        \label{fig:small-silhouette-inds}
        \centering
        \includegraphics[width=\imgwidth]{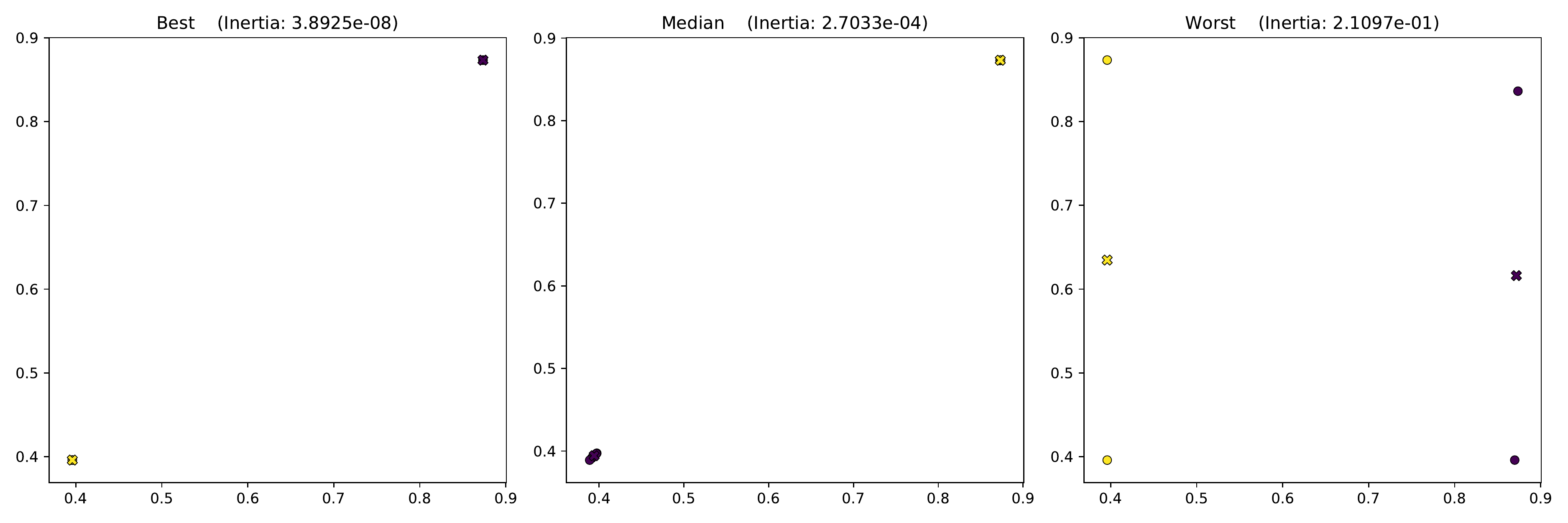}
    }\\

    \subfloat[][]{%
        \label{fig:large-silhouette-inds}
        \centering
        \includegraphics[width=\imgwidth]{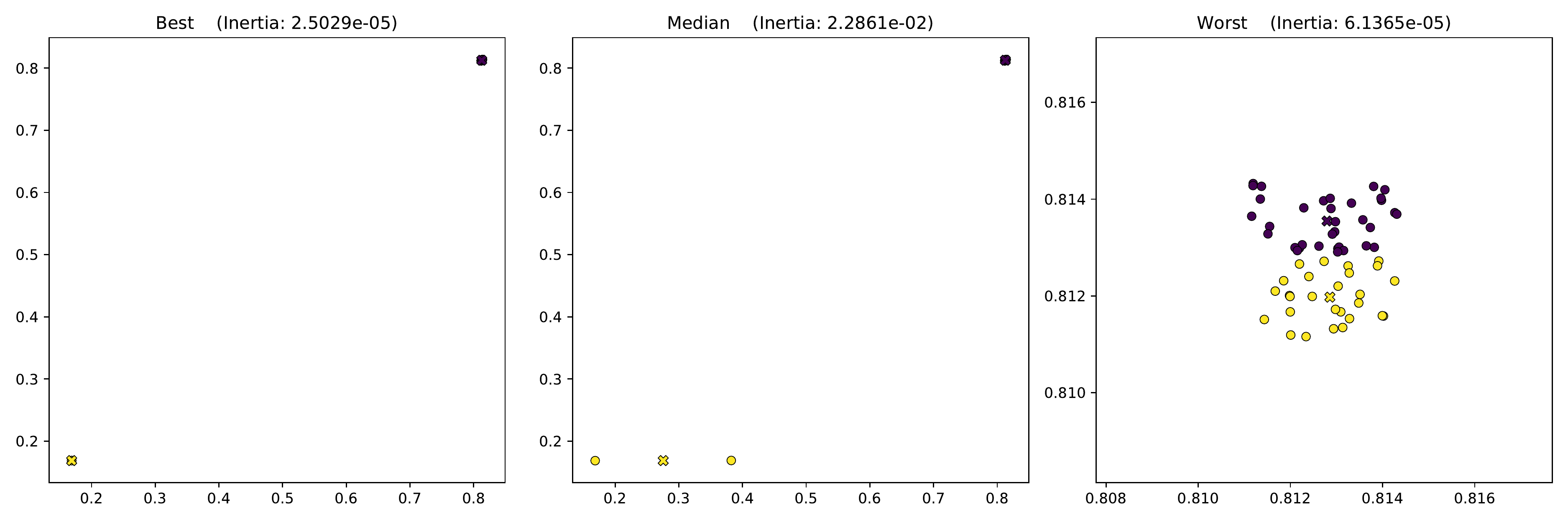}
    }
    \caption[]{%
        Representative individuals based on silhouette with:
        \subref{fig:small-silhouette-inds} \(R~=~(3,100)\);
        \subref{fig:large-silhouette-inds} \(R~=~(50,100)\). Centroids displayed
        as crosses.
    }\label{fig:silhouette-inds}
\end{figure}

\subsection{Comparison with DBSCAN}\label{subsec:dbscan}

The extent of the capabilities EDO holds as a tool to better understand an
algorithm are especially apparent when comparing an algorithm against another
(or set of others) simultaneously. This is done by utilising the freedom of
choice in a fitness function for EDO.\ Consider two algorithms, \(A\) and \(B\),
and some common metric between them, \(g\). Then their similarities and
contrasts can be explored by considering the differences in this metric on the
two algorithms. In terms of EDO, this means using \(f = g_A - g_B\), \(f = g_B -
g_A\) or \(f = \left| g_B - g_A \right|\) as the fitness function. By doing so,
pitfalls, edge cases or fundamental conditions for the method may be
highlighted. Overall, this process allows the researcher to more deeply learn
about the method of interest beyond the traditional method of literature
comparison on a particular example.

Consider the following use case with another clustering algorithm of a different
form, Density Based Spatial Clustering of Applications with Noise (DBSCAN). In
this particular case, the objective is to find datasets for which the method of
interest, \(k\)-means, outperforms its alternative, DBSCAN.\ Here there is no
concept of inertia as DBSCAN is density-based and is able to identify
outliers~\cite{Ester1996}. As such, a valid metric must be chosen. One such
metric is the silhouette score as defined in~(\ref{eq:silhouette}).

In this case, however, an adjustment to the fitness function must be made so as
to accommodate for the condition of the silhouette coefficient that there must
be more than one cluster present. Let \(S_k (X)\) and \(S_D (X)\) denote the
silhouette coefficients of the clustering found by \(k\)-means and DBSCAN
respectively. Then the fitness function is defined to be:
\begin{equation}
    f(X) = 
        \begin{cases}
            S_D (X) - S_k (X), &\quad \text{%
                \begin{tabular}{l}%
                    if DBSCAN identifies two or
                    \\
                    more clusters (inc.\ noise)
                \end{tabular}
            }\\
            \infty &\quad \ \ \text{otherwise.}
        \end{cases}\label{eq:dbscan-fitness}
\end{equation}

There are several remarks to be made here. First, note the order of the
subtraction here as EDO minimises fitness functions by default. Also, \(f\)
takes values in the range \([-2, 2]\) where \(-2\) is the best, i.e.\ \(S_D(X) =
-1\) and \(S_k(X) = 1\). Likewise, 2 is the worst score. Finally, the silhouette
coefficient requires at least two clusters to be present and so if DBSCAN
identifies a single cluster then that individual will be penalised heavily under
this fitness function when, in fact, that clustering may be of high quality. As
such, this fitness function may require adjustment.

It must also be acknowledged that \(k\)-means and DBSCAN share no common
parameters and so direct comparison is more difficult. For the purposes of this
example, only one set of parameters is used but a thorough investigation should
include a parameter sweep in similar, real-world use cases. The parameters being
used are \(k~=~3\) for \(k\)-means, and \(\epsilon~=~0.1,\ MinPoints~=~5\) for
DBSCAN.\ This set was chosen following informal experimentation using the Python
library Scikit-learn~\cite{scikit} to find comparable parameters in the given
search space defined by the EDO parameters used previously with
\(R~=~(50,100)\).

\begin{figure}[htbp]
    \centering
    \begin{tabular}{c}
        \includegraphics[width=\imgwidth]{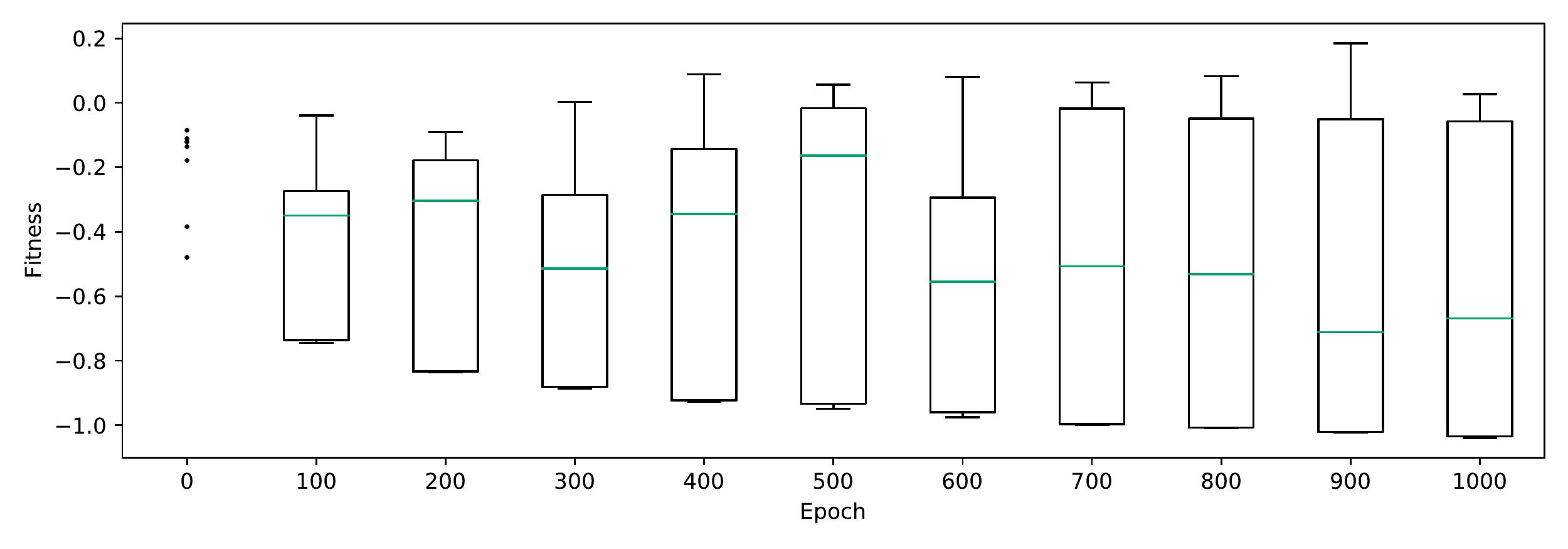}
        \\
        \includegraphics[width=\imgwidth]{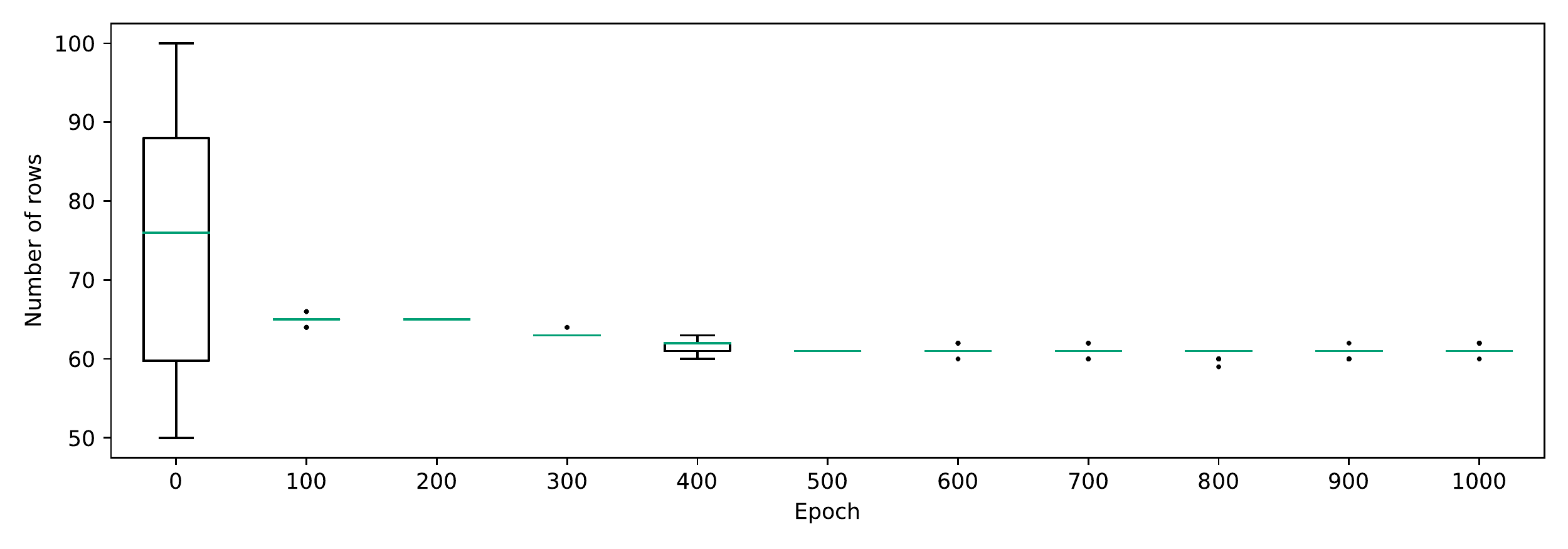}
    \end{tabular}
    \caption{%
        Progressions for difference in silhouette (\(k\)-means-preferable) and
        dimension across 1000 epochs at 100 epoch intervals.
    }\label{fig:dbscan-silhouette}
\end{figure}

\begin{figure}[htbp]
    \centering
    \subfloat[][]{%
        \label{fig:dbscan-inds-k}
        \centering
        \includegraphics[width=\imgwidth]{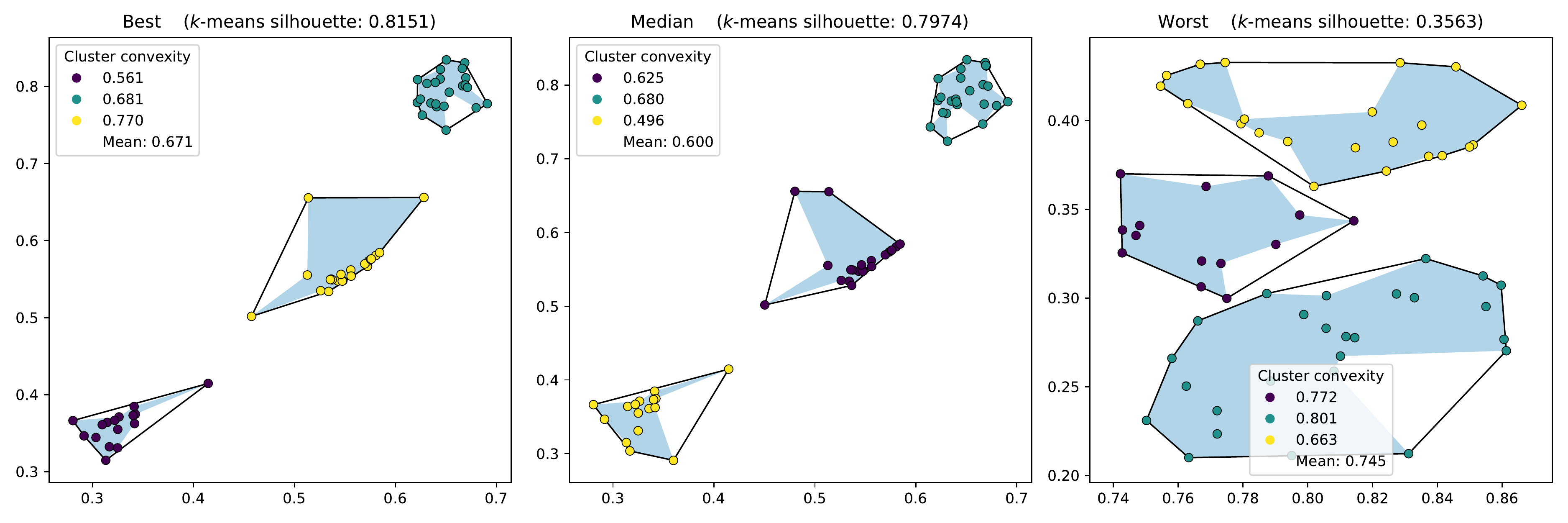}
    }\\
    \subfloat[][]{%
        \label{fig:dbscan-inds-d}
        \centering
        \includegraphics[width=\imgwidth]{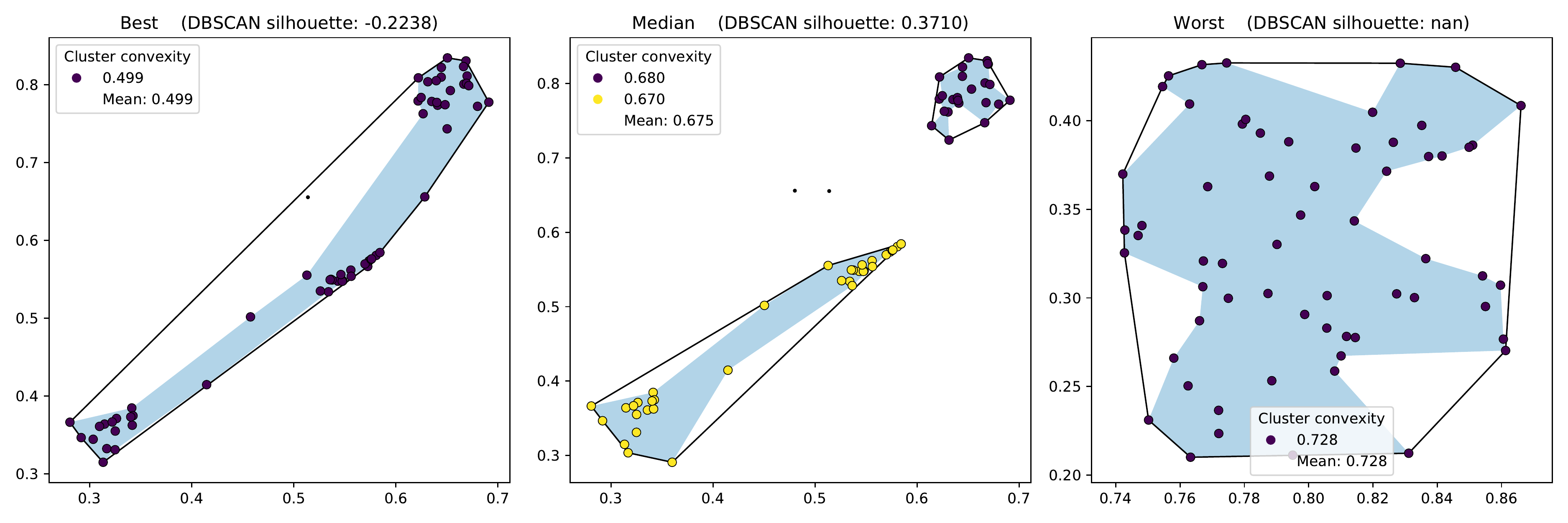}
    }
    \caption[]{%
        Representative individuals from a \(k\)-means-preferable run with
        clustering by: \subref{fig:dbscan-inds-k} \(k\)-means;
        \subref{fig:dbscan-inds-d} DBSCAN.\ Concave and convex hulls illustrated
        by shading and outline respectively. 
    }\label{fig:dbscan-inds}
\end{figure}

Figure~\ref{fig:dbscan-silhouette} shows a summary of the progression of EDO
for this use case. As with the previous examples where \(R~=~(50, 100)\), the
variation in the population fitness is unstable but there is a clear trend of
improvement in the best individual over the course of the run. There is also a
convergence seen in the number of rows a dataset has. The resting dimension
varied across the trials conducted in this work but none exhibited a dramatic
shift toward the lower limit of 50 rows as with previous examples. This is
suggestive of a more competitive environment for individuals where slight
changes to an individual can drastically alter their fitness.

The effect of such changes can be seen in Figure~\ref{fig:dbscan-inds} where
representative individuals are shown for this example. Here, the best performing
individual, when clustered by \(k\)-means, shows three clear and nicely
separated clusters. Note that they are not so tightly packed; again, this
suggests that the route to an optimal individual is less clearly defined. In
contrast, when the same dataset is clustered by DBSCAN a single cluster is found
with a single noise point held within the convex hull of the cluster, i.e.\
there are overlapping clusters (since noise points form a single cluster).
Hence, along with the fact that the larger cluster is widely spread, it follows
that the clustering has a relatively small, negative silhouette coefficient.

Another point of interest here is the convexity of the clusters. A known
condition for the success of \(k\)-means is that the presented clusters are of
roughly equal size and are convex. This is due to the overall objective being to
approximate the centroidal Voronoi tessellation~\cite{Du2006}. Without this
condition, up to the correct choice of \(k\), the algorithm will fail to produce
adequate results for either inertia or silhouette. DBSCAN, on the other hand,
does not have this condition and is able to detect non-convex clusters so long
as they are dense enough. Figure~\ref{fig:dbscan-inds} shows the clustering
found by each method and the respective convex and concave hulls of the clusters
found. The `concave hull' of a cluster is taken to be the \(\alpha\)-shape of
the cluster's data points~\cite{Edelsbrunner1983} where \(\alpha\) is determined
to be the smallest value such that all the points in the cluster are contained
in a single polygon. The convexity of cluster \(Z_j\), denoted
\(\mathcal{C}_j\), is then determined to be the ratio of the area of its concave
hull, \(H_c\), to the area of its convex hull, \(H_v\)~\cite{Sonka1993}:
\begin{equation}
    \mathcal{C}_j := \frac{area(H_c)}{area(H_v)}
\end{equation}

With this definition, it should be clear that a perfectly convex cluster, such
as a single point or line, would have \(\mathcal{C}_j = 1\).

It can be seen that the convexity of the clustering found by \(k\)-means appears
to be higher than that by DBSCAN.\ This was apparent across all trials conducted
in this work and indicates that the condition for convex clusters is being
sought out during the optimisation process. Meanwhile, however, it is not clear
whether the performance of DBSCAN falls owing to its parameters or the method
itself. This is a point where parameter sweeping would prove most useful so as
to determine a crossing point for these two driving forces.

Now, to add to the discussion above, the inverse optimisation should be
considered. That is, using the same parameters, the datasets for which DBSCAN
outperforms \(k\)-means with respect to the silhouette coefficient are to be
investigated. This is equivalent to using \(-f\) as the fitness function
except with the same penalty of \(\infty\) for the case set out
in~(\ref{eq:dbscan-fitness}).

\begin{figure}[htbp]
    \centering
    \begin{tabular}{c}
        \includegraphics[width=\imgwidth]{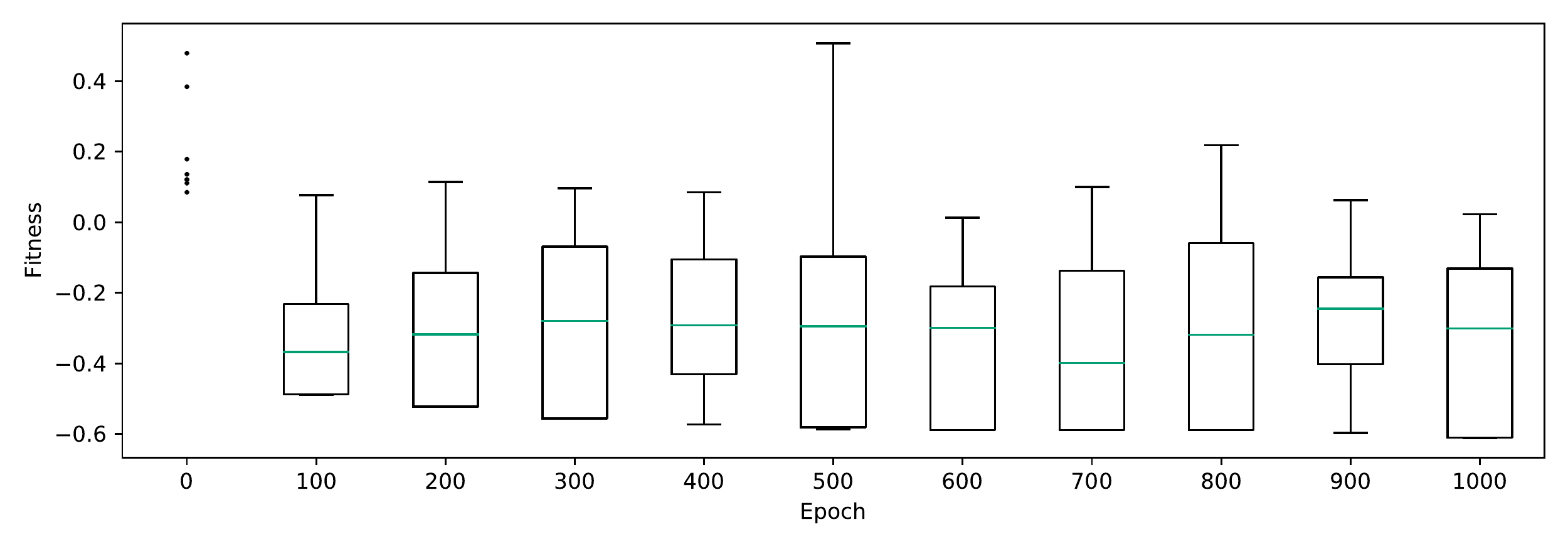}
        \\
        \includegraphics[width=\imgwidth]{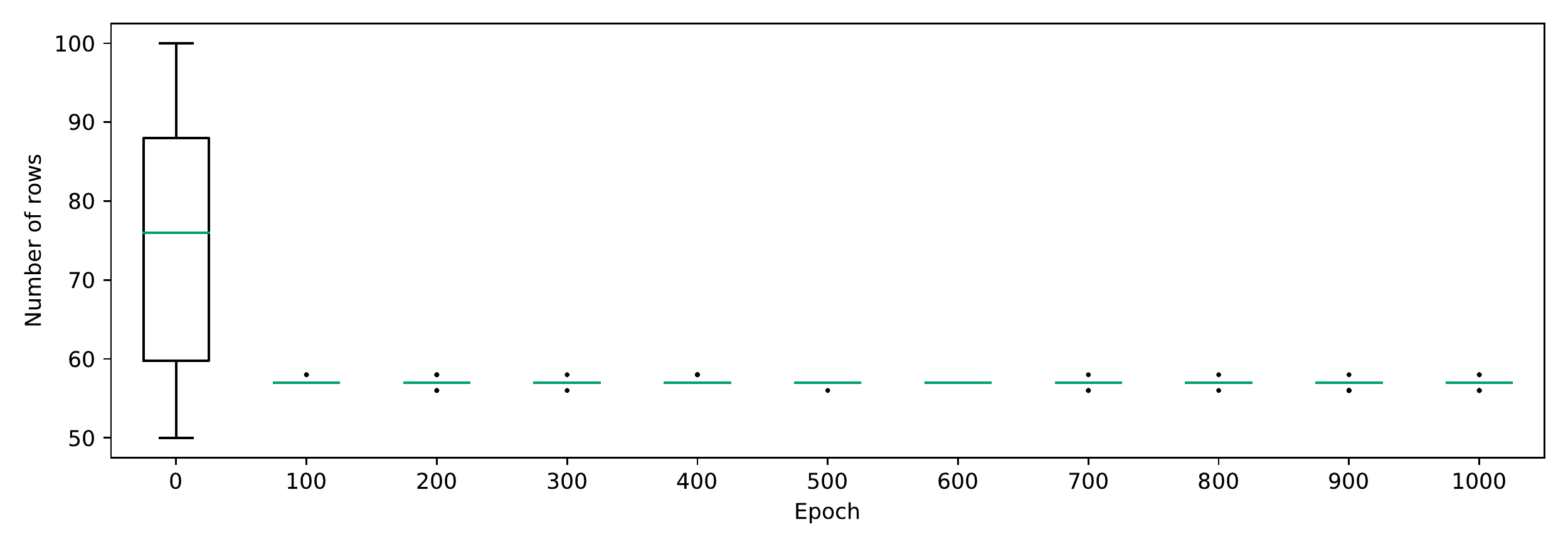}
    \end{tabular}
    \caption{%
        Progressions for difference in silhouette (DBSCAN-preferable) and
        dimension across 1000 epochs at 100 epoch intervals.
    }\label{fig:negative-prog}
\end{figure}

\begin{figure}[htbp]
    \centering
    \subfloat[][]{%
        \label{fig:neg-inds-k}
        \centering
        \includegraphics[width=\imgwidth]{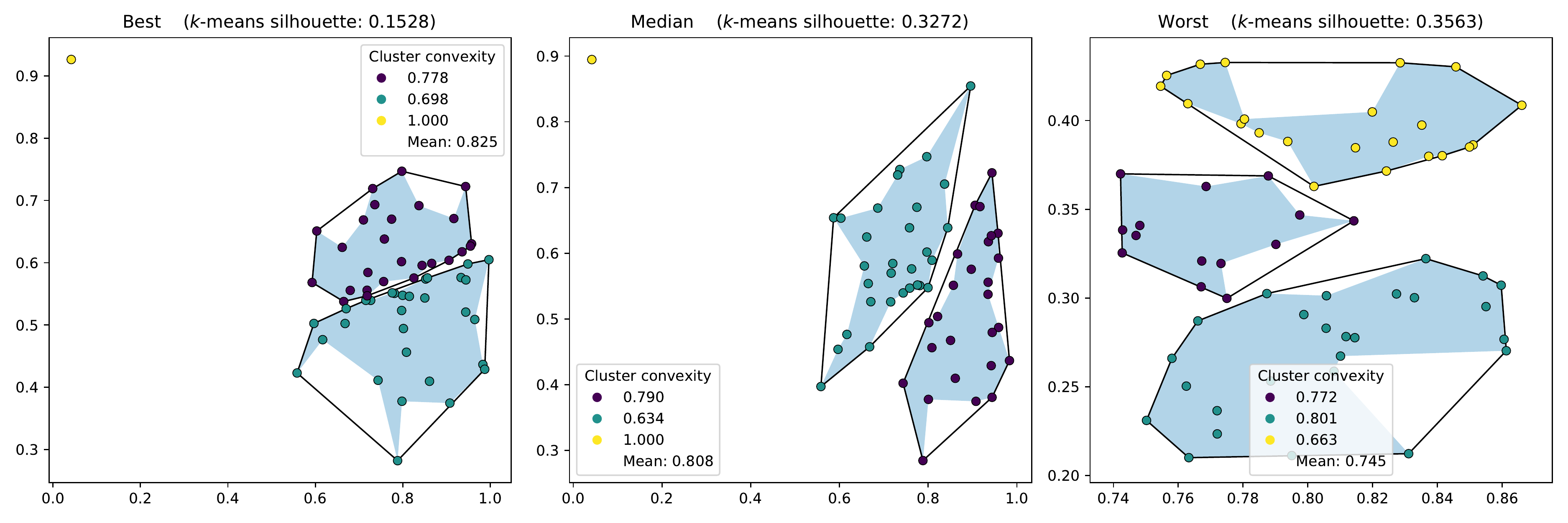}
    }\\
    \subfloat[][]{%
        \label{fig:neg-inds-d}
        \centering
        \includegraphics[width=\imgwidth]{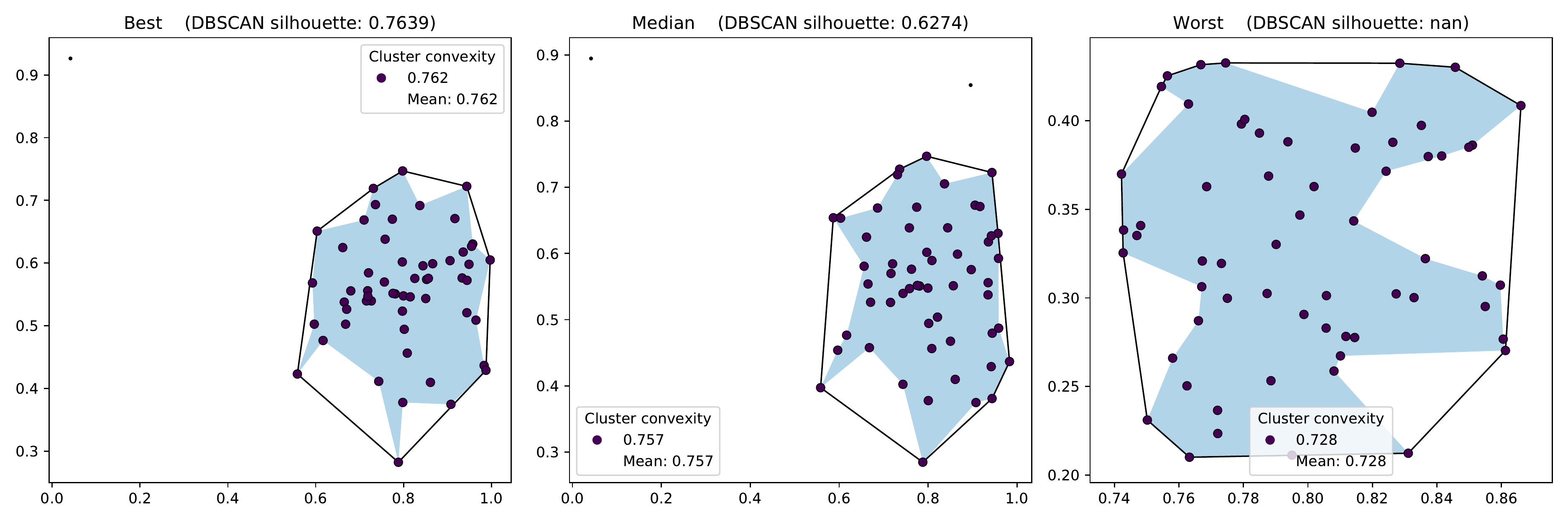}
    }
    \caption[]{%
        Representative individuals from a DBSCAN-preferable run with clustering
        by: \subref{fig:dbscan-inds-k} \(k\)-means; \subref{fig:dbscan-inds-d}
        DBSCAN.\ Concave and convex hulls illustrated by shading and outline
        respectively. 
    }\label{fig:negative-inds}
\end{figure}

Figures~\ref{fig:negative-prog}~and~\ref{fig:negative-inds} show the same
summary as above with the revised fitness function. Inspecting the former, it is
seen that the best fitness found is worse than with the previous example. This,
in part, is due to the fact that \(k\)-means cannot find a clustering with
negative values as no clusters may overlap. It can, however, produce results
with small silhouette scores where the clusters are tightly packed. Hence, the
best fitness score is now \(-1\) whereas the worst is 2, still.

Note in the first two frames of Figure~\ref{fig:neg-inds-k} how \(k\)-means is
forced to split what is evidently a single cluster in two whereas DBSCAN is able
to identify the single cluster and the outlying noise
(Figure~\ref{fig:neg-inds-d}). The proximity of these clusters has then dragged
the silhouette score down for \(k\)-means. Referring to
Figure~\ref{fig:neg-inds-d}, this kind of behaviour is certainly preferable for
DBSCAN under these parameters: the beginning individuals are likely random
clouds (as seen in the rightmost two frames of the figure) and the simplest step
toward a fit dataset is one that maintains that vaguely dense body with minimal
noise points far from it.

As has already been stated, the software implementation of the EDO method
has been produced in line with the best practices of open source software
development and reproducible research. In aid of this, all of the source code
used in these examples (including to create the figures) has been archived
under the DOI
\href{https://doi.org/10.5281/zenodo.3492236}{10.5281/zenodo.3492236}.
Likewise, all of the data produced to support this case study have been archived
under the DOI
\href{https://doi.org/10.5281/zenodo.3492228}{10.5281/zenodo.3492228}.

\section{Conclusion}

In this paper we have introduced a novel approach to understanding the quality
of an algorithm by exploring the space in which their well-performing datasets
exist. Following a detailed explanation of its internal mechanisms, a case study
in \(k\)-means clustering was offered as validation for the proposed method.
The method was able to reveal some known results without prior knowledge when
investigating \(k\)-means in several scenarios, and again when comparing
\(k\)-means and another leading clustering method, DBSCAN.\

The method itself utilises biological operators to traverse a potentially broad
region of the space of all possible datasets. This is done in an organic way
with a minimal external framework attached. The generative nature of the
proposed method also provides transparency and richness to the solution when
compared to other contemporary techniques for artificial data generation as the
entire history of individuals is preserved. While other search and optimisation
methods exist, the decision to use an EA here was down to this transparency and
the ease with which to implement biological operators that are both meaningful
and easily understood. 

The Evolutionary Dataset Optimisation method is dependent on a number of
parameters set out in this work one of which is the choice of distribution
families, \(\mathcal{P}\); these families go on to define the general
statistical shape of the columns of the datasets that are produced and also
control the present data types. The relationship between columns and their
associated distribution is not causal and appropriate methods should be employed
to understand the structure and characteristics of the data produced before
formal conclusions are made as set out in the case study provided.

\begin{figure}[htbp]
    \includegraphics[width=\imgwidth]{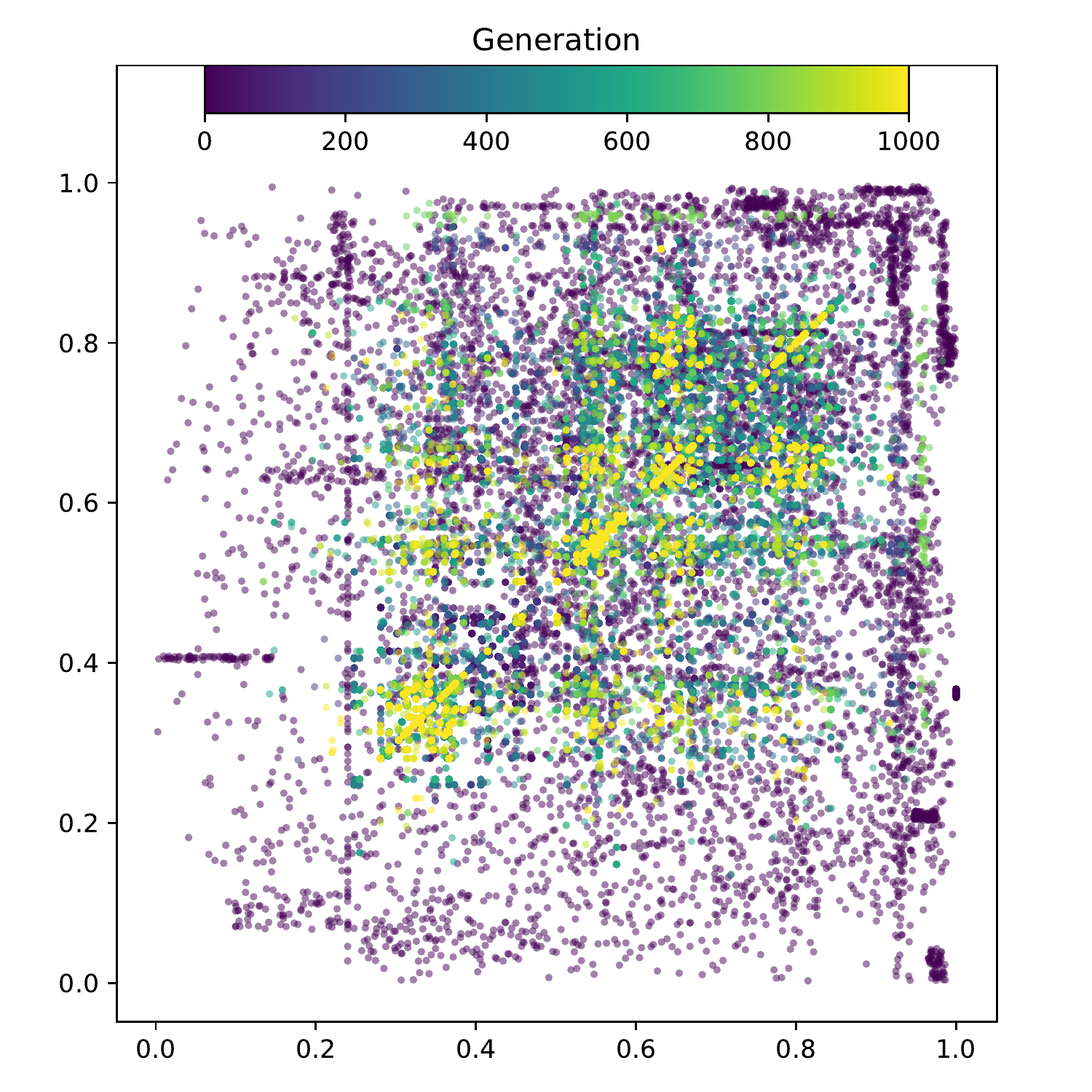}
    \caption{%
        A scatter of all the individuals found at 50 epoch intervals in the
        first example of Section~\ref{subsec:dbscan}, i.e.\ those summarised in
        Figure~\ref{fig:dbscan-inds}.
    }\label{fig:coverage}
\end{figure}

It is known that EAs might terminate at a local optimum and may not be able to
traverse the entire sample space~\cite{Vikhar2016}, or even a sufficient part of
it. This would be even more problematic in the case presented in this work where
the sample space is not even of a fixed size or data type. In all experiments
carried out for this work, this theoretic limitation has not arisen.
Figure~\ref{fig:coverage} shows an exploration of the sample space and it is
evident that the EDO method was able to explore a large proportion of it. In the
early stages, it is also clear here how the EA got stuck in small parts of the
search space before later moving toward a subregion of the unit square.

Although this does provide evidence to say that the EA's current design can
sufficiently explore its given search space, it does not provide any guarantee
that this will happen, even in expectation. Proving this theoretically is an
area for further investigation.

Something that does stand against EAs is their tendancy to find the `easy' way
out. That is, reducing down to the simplest solution which solves the given
problem. In most cases, that is not a problem and is often, in fact, favourable.
Throughout the case study provided, this is seen to happen.
Figure~\ref{fig:coverage} shows this behaviour again by the strong diagonal
region in later generations. In that particular example, the easiest solution
for the EA (i.e.\ for \(k\)-means to outperform DBSCAN) was to collapse one
dimension of the search space to make the problem one-dimensional. This kind of
behaviour is not necessarily a bad thing as trivial, basic and simple cases are
of great importance when understanding an algorithm's quality.

However, should that be a problem, then the objective function could be adjusted
accordingly. In the case study, several iterations of fitness functions were
examined but each was adjusted by hand according to what was apparent at the
time. Due to the architecture of the implementation of this method, this could
be done in practicality. For instance, a similar strategy could be employed
automatically by a more sophisticated fitness function that retains some
information about the datasets generated from previous runs of EDO on a
particular (or at least similar) parameter set. In this way, the currently
completely unsupervised learning conducted by the EA could be ushered away from
less helpful solutions (via some penalty, say) and toward previously unexplored
behaviours. This automatic, iterative application of the proposed method would
likely reveal more sophisticated insights into a particular algorithm.

In essence, the proposed method is merely a tool that demonstrates the benefit
of the flipped paradigm set out in this work. The concept of where `good'
datasets exist is not something that is well-documented in literature and the
hope of this work is that Evolutionary Dataset Optimisation acts as a starting
point for further works to come.


\section*{Acknowledgements}

The authors wish to thank the Cwm Taf Morgannwg University Health Board for
their funding and support of the Ph.D. of which this work has formed a part.

\section*{Conflict of interest}

The authors declare that they have no conflict of interest.

\bibliography{references}

\pagebreak%

\appendix\section{Appendix}

\subsection{Lloyd's algorithm}\label{app:kmeans}

\balg[H]%
\KwIn{a dataset \(X\), a number of centroids \(k\), a distance metric \(d\)}
\KwOut{a partition of \(X\) into \(k\) parts, \(Z\)}

\Begin{%
    select \(k\) initial centroids, \(z_1, \ldots, z_k \in X\)\;
    \While{any point changes cluster or some stopping criterion is not met}{%
        assign each point, \(x \in X\), to cluster \(Z_{j^*}\) where:
        \[
            j^* = \argmin_{j = 1, \ldots, k} \left\{%
                {d\left(x, z_j\right)}^2
            \right\}
        \]\;
        recalculate all centroids by taking the intra-cluster mean:
        \[
            z_j = \frac{1}{|Z_j|} \sum_{x \in Z_j} x
        \]
    }
}
\caption{\(k\)-means (Lloyd's)}
\ealg%

\subsection{Implementation example}\label{app:code}

Below is an example of how the Python implementation was used to complete the
first example, including the definition of the fitness function.

\begin{lstlisting}
import edo
from edo.pdfs import Uniform
from sklearn.cluster import KMeans

def fitness(dataframe, seed):
    """ Return the final inertia of 2-means on the `dataframe`
    for the given `seed`. """

    km = KMeans(n_clusters=2, random_state=seed).fit(dataframe)
    return km.inertia_

Uniform.param_limits["bounds"] = [0, 1]

pop_history, fit_history = edo.run_algorithm(
    fitness, size=100, row_limits=[3, 100], col_limits=[2, 2],
    families=[Uniform], max_iter=1000, best_prop=0.2,
    mutation_prob=0.01, seed=0, root="out",
    fitness_kwargs={"seed": 0},
)
\end{lstlisting}

\end{document}